\documentclass[useAMS,usegraphicx]{mn2e}
\title[UVES and FORS2 spectroscopy of GRB081008]{VLT/UVES
  and FORS2 spectroscopy of the GRB\,081008 afterglow\thanks{Based on
    observations collected at the European Southern Observatory, ESO,
    the VLT/Kueyen telescope, Paranal, Chile, in the framework of the
    program 082-0755.}}

\author[V. D'Elia et al.]{V. D'Elia$^{1,2}$\thanks{E-mail:
delia@asdc.asi.it}, S. Campana$^3$,
S. Covino$^3$, P. D'Avanzo$^3$, S. Piranomonte$^{1}$, G. Tagliaferri$^3$ 
\\
$^1$ INAF-Osservatorio Astronomico di Roma, Via Frascati 33, I-00040 Monteporzio Catone, Italy\\ 
$^2$ ASI-Science Data Centre, Via Galileo Galilei, I-00044 Frascati, Italy\\ 
$^3$ INAF, Osservatorio Astronomico di Brera, Via E. Bianchi 46, 23807 Merate (LC), Italy\\ 
}
\begin{document}

\date{Accepted... Received...; in original form...}

\pagerange{\pageref{firstpage}--\pageref{lastpage}} \pubyear{2002}

\maketitle

\label{firstpage}

\begin{abstract}

  We aim at studying the gamma-ray burst GRB\,081008 environment by
  analysing the spectra of its optical afterglow.  UVES/VLT
  high resolution spectroscopy of GRB\,081008 was secured $\sim 5$ hr
  after the {\it Swift}-BAT trigger. 
Our dataset comprises also three
  VLT/FORS2 
  nearly simultaneous spectra of the same source. The availability of
  nearly simultaneous high and low resolution spectra for a GRB
  afterglow is an extremely rare event. The GRB-Damped Lyman Alpha
  system at $z = 1.9683$ shows that the interstellar medium (ISM) of
  the host galaxy is constituted by at least three components which
  contribute to the line profiles. Component I is the redmost one, and
  is $20$ km/s and $78$ km/s redward component II and III,
  respectively. We detect several ground state and
  excited absorption features in components I and II. These features
  have been used to compute the distances between the GRB and the
  absorbers. Component I is found to be $52 \pm 6$ pc away from the
  GRB, while component II presents few excited transitions and its
  distance is $200^{+60}_{-80}$ pc. Component III only features a few,
  low ionization and saturated lines suggesting that it is even
  farther from the GRB. Component I represents the closest absorber
  ever detected near a GRB. This (relatively) low distance can
  possibly be a consequence of a dense GRB environment, which prevents
  the GRB prompt/afterglow emission to strongly affect the ISM up to
  higher distances.  The hydrogen column density associated to
  GRB\,081008 is $\log N_{\rm H}/{\rm cm}^{-2} = 21.11 \pm 0.10$, and
  the metallicity of the host galaxy is in the range [X/H] $= -1.29$
  to $-0.52$. In particular, we found [Fe/H]$=-1.19 \pm 0.11$ and
  [Zn/H]$=-0.52\pm 0.11$ with respect to solar values. This
  discrepancy can be explained by the presence of dust in the GRB ISM,
  given the opposite refractory properties of iron and zinc. By
  deriving the depletion pattern for GRB\,081008, we find the optical
  extinction in the visual band to be $A_V \sim 0.19$ mag. The Curve
  of Growth analysis applied to the FORS2 spectra brings column
  densities consistent at the $3\sigma$ level to that evaluated from
  the UVES data using the line fitting procedure. This reflects the
  low saturation of the detected GRB\,081008 absorption features.

\end{abstract}


\begin{table*}
\begin{center}
\caption{UVES and FORS2 setups.}
{\footnotesize
\smallskip
\begin{tabular}{|l|c|c|c|c|c|c|c|}
\hline 
Instrument &Setup (nm) & Time from burst (hr) & Exposure (s)       & Wavelength (\AA)&  Slit width & Resolution        &  S/N        \\
\hline 
UVES       &Dic 1, 346 & $4.30$               & $1800$             & 3300 - 3870     &  1''        & 40 000            &  $\sim 3-5$    \\
\hline
UVES       &Dic 1, 580 & $4.30$               & $1800$             & 4780 - 6810     &  1''        & 40 000            &  $\sim 5-8$    \\
\hline
FORS2      &600B+22(A) & $4.37$               &  $900$             & 3300 - 6300     &  1''        &    780            &  $\sim 35 - 50$     \\
\hline 
FORS2      &600B+22(B) & $4.63$               &  $900$             & 3300 - 6300     &  1''        &    780            &  $\sim 35 - 50$     \\
\hline
FORS2      &600B+22(C) & $4.88$               &  $900$             & 3300 - 6300     &  1''        &    780            &  $\sim 35 - 50$     \\
\hline
FORS2      &A+B+C      & $4.63$               & $2700$             & 3300 - 6300     &  1''        &    780            &  $\sim 60 - 80$     \\
\hline
\end{tabular}
}
\end{center}
\end{table*}

\section{Introduction}

Long gamma-ray bursts (GRBs) are powerful explosions occurring at
cosmological distances, linked to the death of massive stars. The
gamma-ray (or prompt) event is followed by an afterglow at longer
wavelengths, which is crucial in order to understand the physics of
these sources, but also to investigate the nature of the interstellar
medium (ISM) of high redshift galaxies. Before GRBs, such studies made
use of Lyman-break galaxies (LBGs, see e.g. Steidel et al.  1999) and
galaxies that happen to be along the lines of sight to bright
background quasars, commonly referred to as QSO-Damped Lyman Alpha
(DLA) systems. However, both classes are entangled by selection
effects. In fact, LBGs fall in the bright end of the galaxy luminosity
function and may not entirely represent typical high-redshift
galaxies. On the other hand, QSO sightlines preferentially probe
galaxy halos, rather than bulges or discs, for cross-section effects
(Fynbo et al. 2008). Indeed, Savaglio et al. (2004; 2005) studied the
ISM of a sample of faint $K$-band selected galaxies at $1.4 < z <
2.0$, finding {Mg}{II} and {Fe}{II} abundances much higher than in QSO
systems but similar to those in gamma-ray burst hosts. Unfortunately,
these galaxies are too faint to be spectroscopically studied up to
higher redshifts, using 8m class telescopes. In this context, long
GRBs can be used as torchlights to illuminate the high-redshift ISM,
and thus represent an independent tool to study high-redshift
galaxies.

Several papers report a metallic content in GRB host galaxies in the
range $10^{-2} - 1 $ with respect to solar values (see e.g., Fynbo et
al. 2006; Savaglio 2006; Prochaska et al. 2007). The GRB host
metallicity is thus on average higher than in QSO-DLA systems,
supporting the notion that GRBs explode well within their hosts. Since
long GRBs are linked to the death of massive stars, they are though to
originate in molecular clouds. In this scenario, absorption from
ground-state and vibrationally excited levels of H$_2$ and other
molecules is expected, but not observed (Vreeswijk et al. 2004;
Tumlinson et al. 2007). The non-detection of these molecular states
(with the exception of GRB\,080607, see Prochaska et al. 2009; Sheffer
et al. 2009) could be a consequence of the intense UV flux from the
GRB afterglow, which photo-dissociates the molecules. However,
molecular hydrogen is not detected in QSO-DLA either (e.g., Noterdaeme
et al. 2008; Tumlinson et al. 2007), possibly indicating that these
molecules are just hard to see at high redshift.

This is just an example of how a GRB can modify its surrounding
medium. The most impressive manifestation of the transient nature of
GRBs in optical spectroscopy is the detection of strong absorption
features related to the excited levels of the {{O}{I}}, {{Fe}{II}},
{{Ni}{II}}, {{Si}{II}} and {{C}{II}} species and their time
  variability (Vreeswijk et al. 2007). This variability can not be
explained assuming infrared excitation or collisional processes
(Prochaska, Chen \& Bloom 2006; Vreeswijk et al. 2007; D'Elia et
al. 2009a), thus excitation by the intense GRB UV flux is the leading
mechanism to produce these features. In this framework, the
GRB/absorber distance can be evaluated comparing the observed
  ground state and excited level abundances with that predicted by
  time-dependent photo-excitation codes. This distance turns out to
be in the range $\sim 0.1-1$ kpc (Vreeswijk et al. 2007; D'Elia et
al. 2009a,b; Ledoux et al. 2009).

Within the described framework, the best and most complete tool to
perform these kind of studies is high resolution spectroscopy. In
fact, it is the only way to disentangle the GRB interstellar medium in
components and to separate the contribution to the absorption coming
from the excited levels from the ground state ones. In addition, a
high spectral resolution allows us to check for saturation of
lines a few km s$^{-1}$ wide that may appear unsaturated in lower
resolution spectra (see e.g. Penprase et al. 2010).

In this paper we present data on GRB\,081008, observed both in high
and low resolution UVES and FORS2 at the VLT.
The paper is organized as follows. Section $2$ summarizes the
GRB\,081008 detection and observations from the literature; Sect. $3$
presents the UVES observations and data reduction; Sect. $4$ is
devoted to the study of the features from the host galaxy, in
  particular their metallicity and distance from the GRB explosion
  site; Sect. $5$ presents the FORS2 data and makes a comparison with
the UVES ones; finally in Sect. $6$ the results are discussed and
conclusions are drawn. We assume a concordance cosmology with $H_0=70$
km s$^{-1}$ Mpc$^{-1}$, $\Omega_{\rm m} = 0.3$, $\Omega_\Lambda =
0.7$. Hereafter, with [X/H] we refer to the X element abundance
relative to solar values.

\section{GRB\,081008}

GRB\,081008 was discovered by {\it Swift}/BAT on October 8, 2008, at
19:58:29 UT, and was detected by both the XRT and the UVOT instruments
(Racusin et al. 2008). The UVOT magnitude in the white filter was
reported to be $15.0$ at $96$ s from the trigger. The afterglow was
also detected in all filters (from {\it B} to {\it K}) by SMARTS/ANDICAM $\sim 4$
hr post burst (Cobb 2008). The redshift was secured by the
Gemini-South/GMOS, which observed the afterglow $5$ hr after the {\it
  Swift} trigger, reporting a redshift of $z=1.967$ (Cucchiara et
al. 2008a). This value was later confirmed by our VLT/UVES+FORS2 data
(D'Avanzo et al. 2008). The host galaxy was identified in the
Gemini-South/GMOS acquisition image, and spectroscopically confirmed
to be at the GRB redshift. The host of GRB\,081008 has $R=20.75\pm
0.01$, which corresponds to an absolute AB magnitude of $-21.5$
(Cucchiara et al. 2008b). A multi-wavelength study of the prompt event
and the early afterglow phase of GRB\,081008 is reported by Yuan et
al. (2010, hereafter Y10), which present {\it Swift} (BAT+XRT+UVOT),
ROTSE-III and GROND data.

\begin{table}
\begin{center}
\caption{Rest frame equivalent widths of the UVES features.}
{\footnotesize
\smallskip
\begin{tabular}{|l|c|c|c|}
\hline 
Species    &Transition & W$_r$ (\AA)          & $\Delta$ W$_r$ (\AA, $1\sigma$)    \\
\hline 
OI$^3P_{2}$ (g.s)        &$1302$     &$0.57$   &$0.05$ \\
\hline
AlII$^1S_0$ (g.s)        &$1670$     &$0.67$   &$0.01$ \\
\hline
AlIII$^2S_{1/2}$ (g.s)    &$1854$     &$0.22$   &$0.01$ \\
                         &$1862$     &$0.13$   &$0.02$ \\
\hline
SiII$^2P^{0}_{1/2}$ (g.s) &$1260$     &$0.63$   &$0.07$ \\
                         &$1808$     &$0.20$   &$0.02$ \\
\hline
SiII$^2P^{0}_{3/2}$ (1*)  &$1264$     &$0.63$   &$0.07$ \\
                         &$1816$     &$0.06$   &$0.02$ \\
\hline
CrII$^2S_{1/2}$ (g.s.)    &$2056$     &$0.20$   &$0.02$ \\
                         &$2062$     &$0.14$   &$0.02$ \\
                         &$2066$     &$0.11$   &$0.02$ \\
\hline
FeII$a^6D_{9/2}$ (g.s)    &$2249$     &$0.12$   &$0.01$ \\
                         &$2260$     &$0.20$   &$0.02$ \\
\hline
FeII$a^6D_{7/2}$ (1*)     &$1618$     &$0.05$   &$0.01$ \\
                         &$1621$     &$0.11$   &$0.01$ \\
\hline
FeII$a^6D_{5/2}$ (2*)     &$1629$     &$0.03$   &$0.01$ \\
\hline
FeII$a^6D_{3/2}$ (3*)     &$1634$     &$0.03$   &$0.01$ \\
                         &$1636$     &$0.03$   &$0.01$ \\
\hline
FeII5s$a^4F_{9/2}$ (5*)   &$1637$     &$0.04$   &$0.01$ \\
                         &$1612$     &$0.12$   &$0.01$ \\
                         &$1702$     &$0.21$   &$0.02$ \\
\hline
FeII$a^4D_{7/2}$ (9*)    &$1635$     &$0.03$   &$0.01$ \\
\hline
NiII$^2D_{5/2}$  (g.s.)  &$1741$     &$0.07$   &$0.02$ \\
\hline
NiII$^4F_{9/2}$ (2*)     &$2166$     &$0.19$   &$0.01$ \\
                        &$2217$     &$0.27$   &$0.01$ \\
                        &$2223$     &$0.09$   &$0.02$ \\
\hline
ZnII$^2S_{1/2}$ (g.s.)   &$2026$     &$0.19$   &$0.02$ \\
                        &$2062$     &$0.09$   &$0.02$ \\
\hline
\end{tabular}
}
\end{center}
\end{table}

\section{UVES observations and data reduction}

The GRB\,081008 afterglow was observed with the high resolution
UV-visual echelle spectrograph (UVES, Dekker et al. 2000), mounted at
the VLT-UT2 telescope, in the framework of the ESO program
082.A-0755. Observations began on the $9^{th}$ October 2008 at
00:16:43 UT ($\sim 4.25$ hr after the {\it Swift}/BAT trigger), when
the magnitude of the afterglow was $R\sim18.5$. Data were acquired
under good observing conditions, with seeing $\sim 0.7$. Only the
UVES-dichroic-1 (red and blue arm) was used due to observational and
scheduling constraints. The net exposure time of the observation is 30
minutes. The slit width was set to be $1''$ (corresponding to a
resolution of $R=40000$) and the read-out mode was rebinned to 2
$\times$ 2 pixels. The spectral range of our observation is
$\sim$3300\AA\ to $\sim$3870\AA, $\sim$4780\AA\ to $\sim$5750\AA, and
$\sim$5830\AA\ to $\sim$6810\AA. Table 1 makes a summary of our
observations.

\begin{figure*}
\centering
\includegraphics[angle=90,width=18cm]{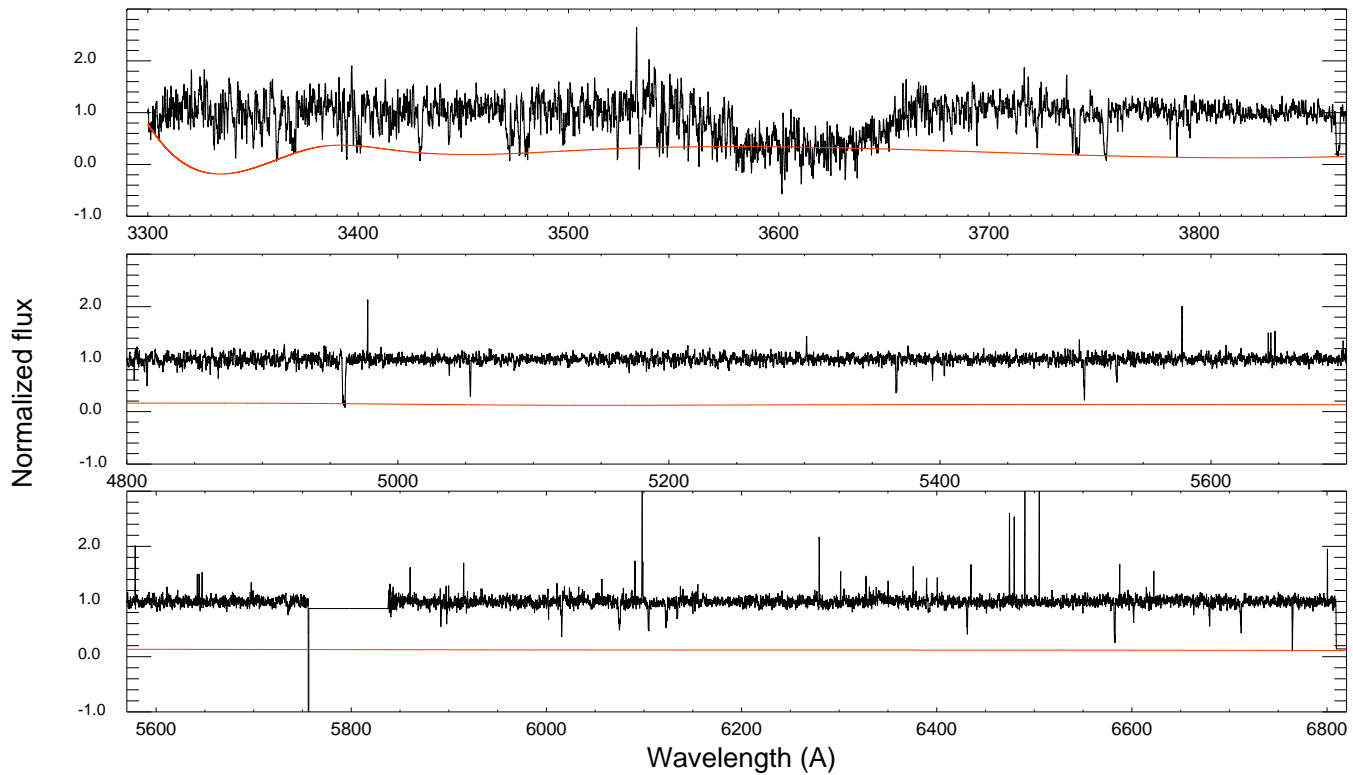}
\caption{The full, {\bf smoothed and} normalized UVES spectrum. Solid lines
  indicate the noise level as a function of wavelength. }
\end{figure*}

The data reduction was performed using the UVES pipeline (version
2.9.7, Ballester et al. 2000). The signal-to-noise ratio per pixel is
$\sim 3-5$ in the blue arm and $\sim 5-8$ in the red one. The noise
spectrum, used to determine the errors in the best-fit line
parameters, was calculated from the real, background-subtracted
spectrum, using line-free regions to evaluate the standard deviation
of continuum pixels. Since the noise spectrum has been produced after
the pipeline processing and the background subtraction, it takes into
account possible systematic errors coming from the data reduction
process. Fig. 1 shows the full, {\bf smoothed and} normalized UVES spectrum.

\begin{figure}
\centering
\includegraphics[angle=-0,width=9cm]{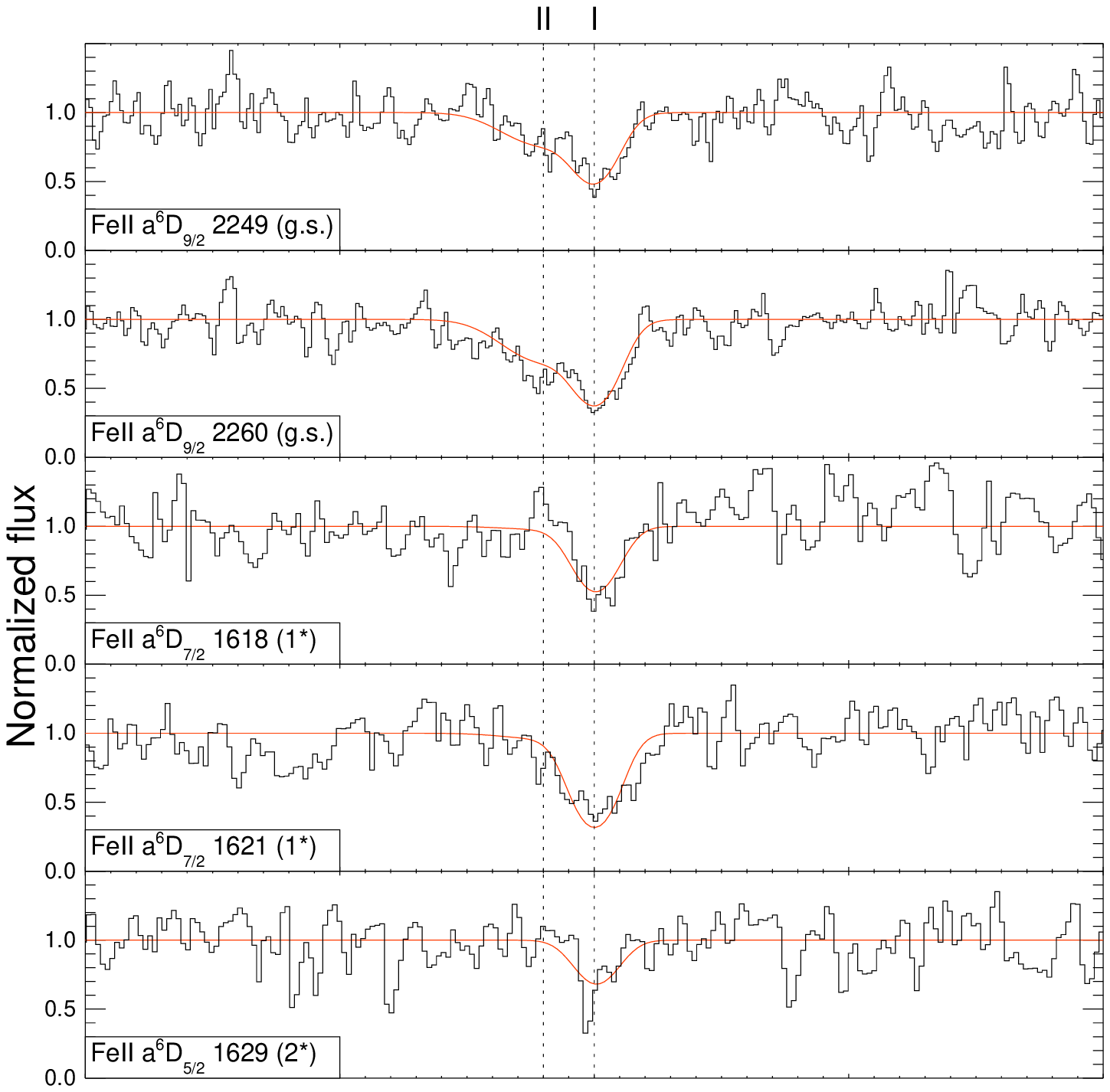}
\includegraphics[angle=-0,width=9cm]{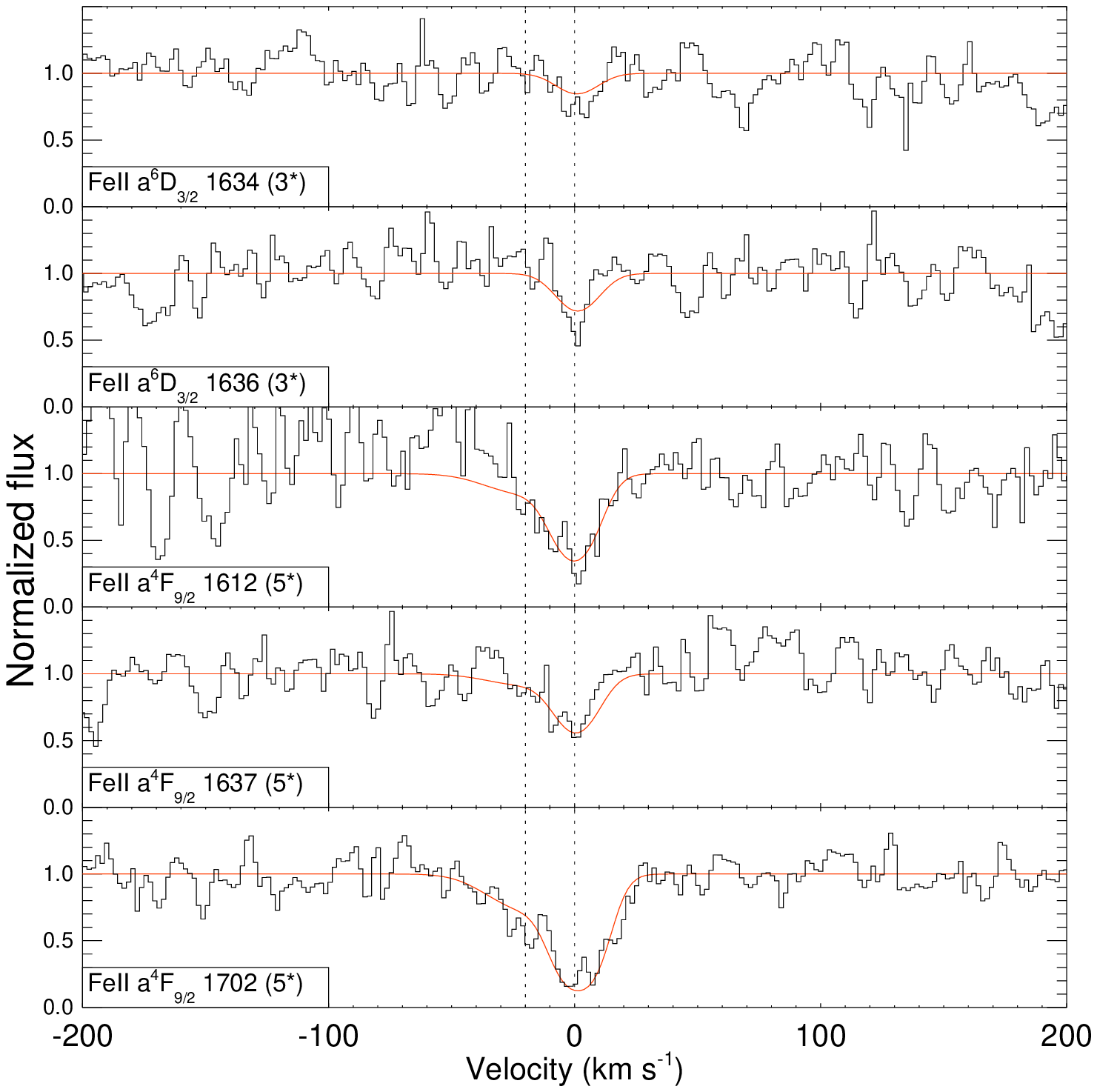}
\caption{The {Fe}{II} ground and excited absorption features. Solid
  lines represent the two Voigt components, best-fit model. Vertical
  lines identify the component velocities. The zero point has been
  arbitrarily placed at the redshift of the redmost component
  ($z=1.9683$). g.s. and n* indicate ground state and n-th excited
  transitions, respectively.}
\label{spe1}
\end{figure}

\section{UVES data analysis}

The gas residing in the GRB host galaxy is responsible for many
features observed in the GRB\,081008 afterglow spectrum.  Metallic
features are apparent from neutral ({{O}{I}}) and low-ionization
({{Al}{II}}, {{Al}{III}}, {{Si}{II}}, {{Cr}{II}}, {{Fe}{II}},
{{Ni}{II}}, {{Zn}{II}}) species. In addition, strong absorption lines
from the fine structure levels of {{Si}{II}}, {{Fe}{II}} and from the
metastable levels of {{Fe}{II}} and {{Ni}{II}} are identified,
suggesting that the intense radiation field from the GRB excites such
features. Table~2 gives a summary of all the absorption lines due to
the host galaxy gas and report their rest frame equivalent widths
  (W$_r$). The spectral features were analyzed with FITLYMAN (Fontana
\& Ballester 1995), using the atomic parameters given in Morton
(2003). The probed ISM of the host galaxy is resolved into two main
components separated by $20$ km s$^{-1}$ (Figs. 2 and 3). The wealth
of metal-line transitions allows us to precisely determine the
redshift of the GRB host galaxy. This yields a vacuum-heliocentric
value of $z=1.9683 \pm 0.0001$, setting the reference point to the
redmost component (hereafter component I). The absorption
  features have been fitted with Voigt profiles, fixing the redshift
  of the two components when studying different lines. All
transitions appear to be nicely lined-up in redshift, with the
exception of component II of {Si}{II}$\lambda$1808. We attribute
  this misalignment to a contamination of another feature, and fit
  just the redmost side of component II.  All ground state and
metastable species present absorption features in both components,
while fine structure levels in component I only.
Two sharp features can be seen at $v=\pm80$ km s$^{-1}$ from the
{Fe}{II} a$^6$D$_{5/2}$ line (Fig. 2). They can not be separated in
the FORS2 spectrum, thus we can not safely assess if they are real or
not.  The Doppler $b$ parameter has been linked between different
excited transitions belonging to the same species.  A small variation
of the Doppler parameter is allowed among different species, but the
fits are quite good even fixing it. The values for components I and II
are $\sim 10$ and $\sim 20$ km s$^{-1}$, respectively. An exception to
this behaviour is represented by component II of {Zn}{II}. In order to
obtain a good fit, a $b \sim 4$ and $\sim 50$ km s$^{-1}$ is required
for component I and II, respectively. This large $b$ value in
component II is necessary to adequately fit what appears as a low
level of the continuum in particular in the {Zn}{II}$\lambda$2026
feature. {\bf The large difference between the b parameters deduced
  for {Zn}{II} and {Cr}{II} is odd, but we do not have a simple
  explanation for it.} The column densities and b parameters for all
the elements and ions of the host galaxy's absorbing gas are reported
in Table 3. A third component is actually identified at $-78$ km
s$^{-1}$ for some low-ionization lines only, i.e., the
{O}{I}$\lambda$1302, {Al}{II}$\lambda$1670 and {Si}{II}$\lambda$1260,
and for the fine structure level {Si}{II}$\lambda$1264 (see
Fig. 4). These lines are however heavily saturated, and their column
densities reported in the table just set a lower limit to the true
values. The reported upper limits are at the 90\% confidence level.

\begin{figure}
\centering
\includegraphics[angle=-0,width=8.5cm,height=7cm]{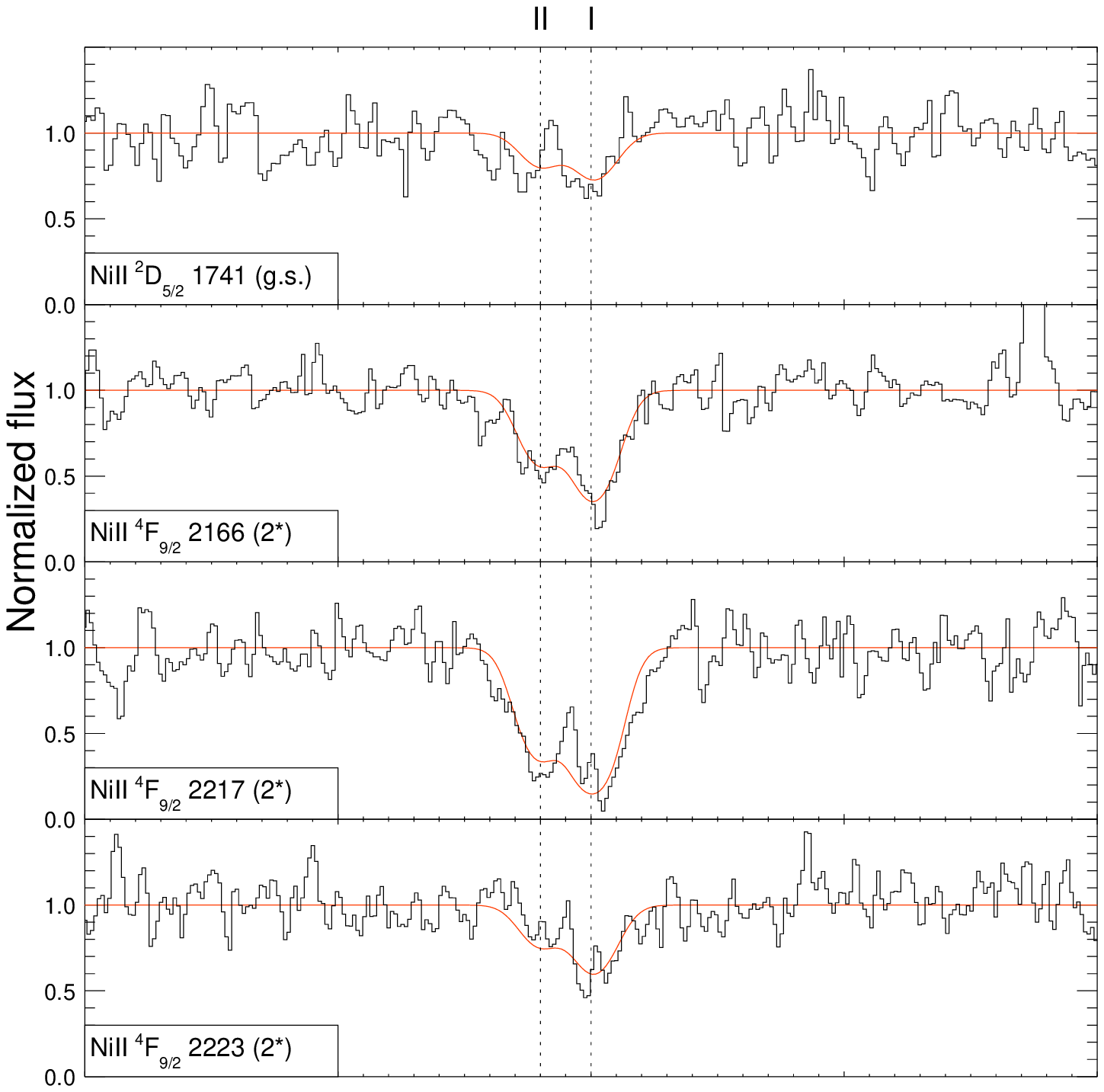}
\includegraphics[angle=-0,width=8.5cm,height=7cm]{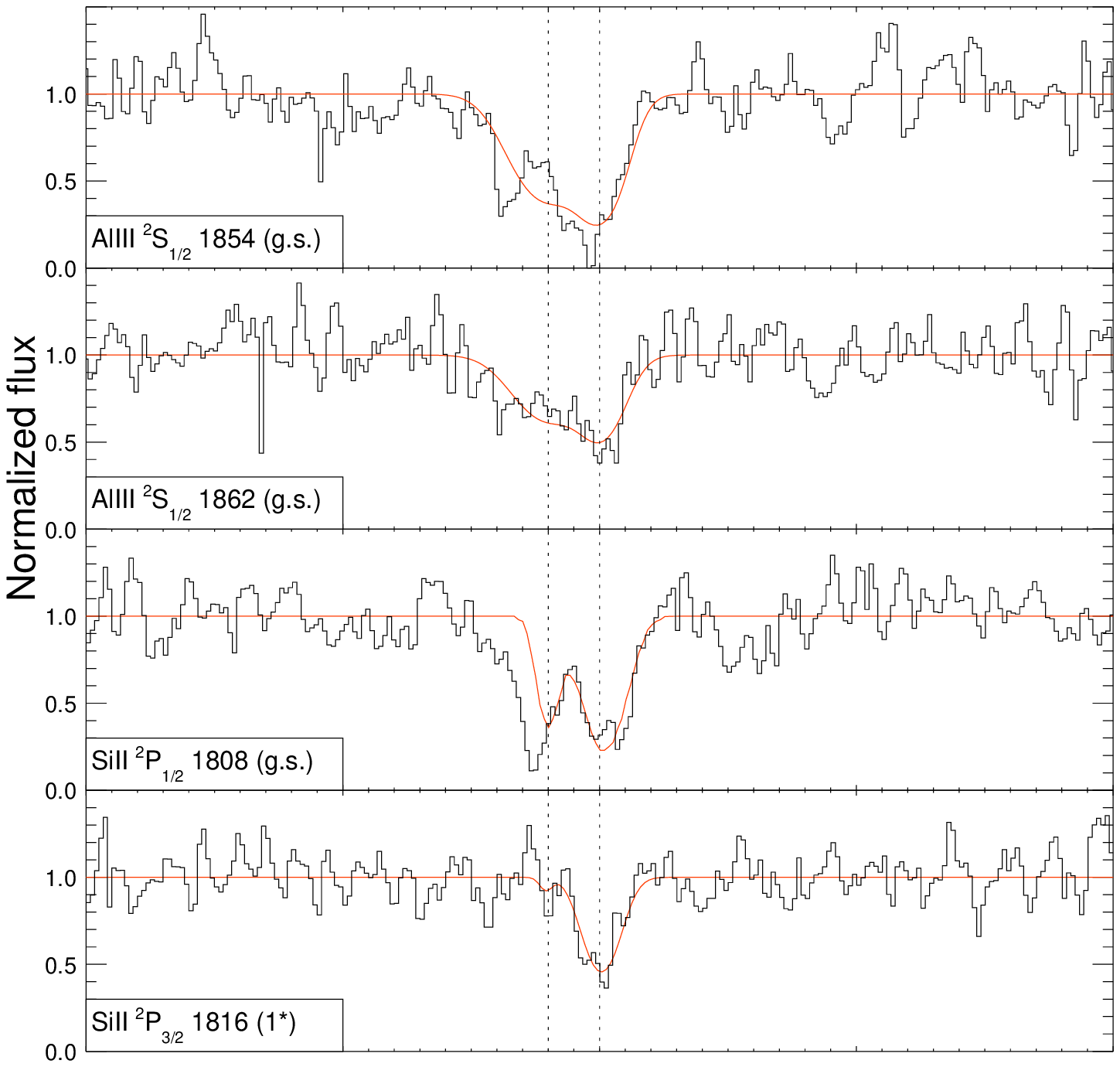}
\includegraphics[angle=-0,width=8.5cm,height=7cm]{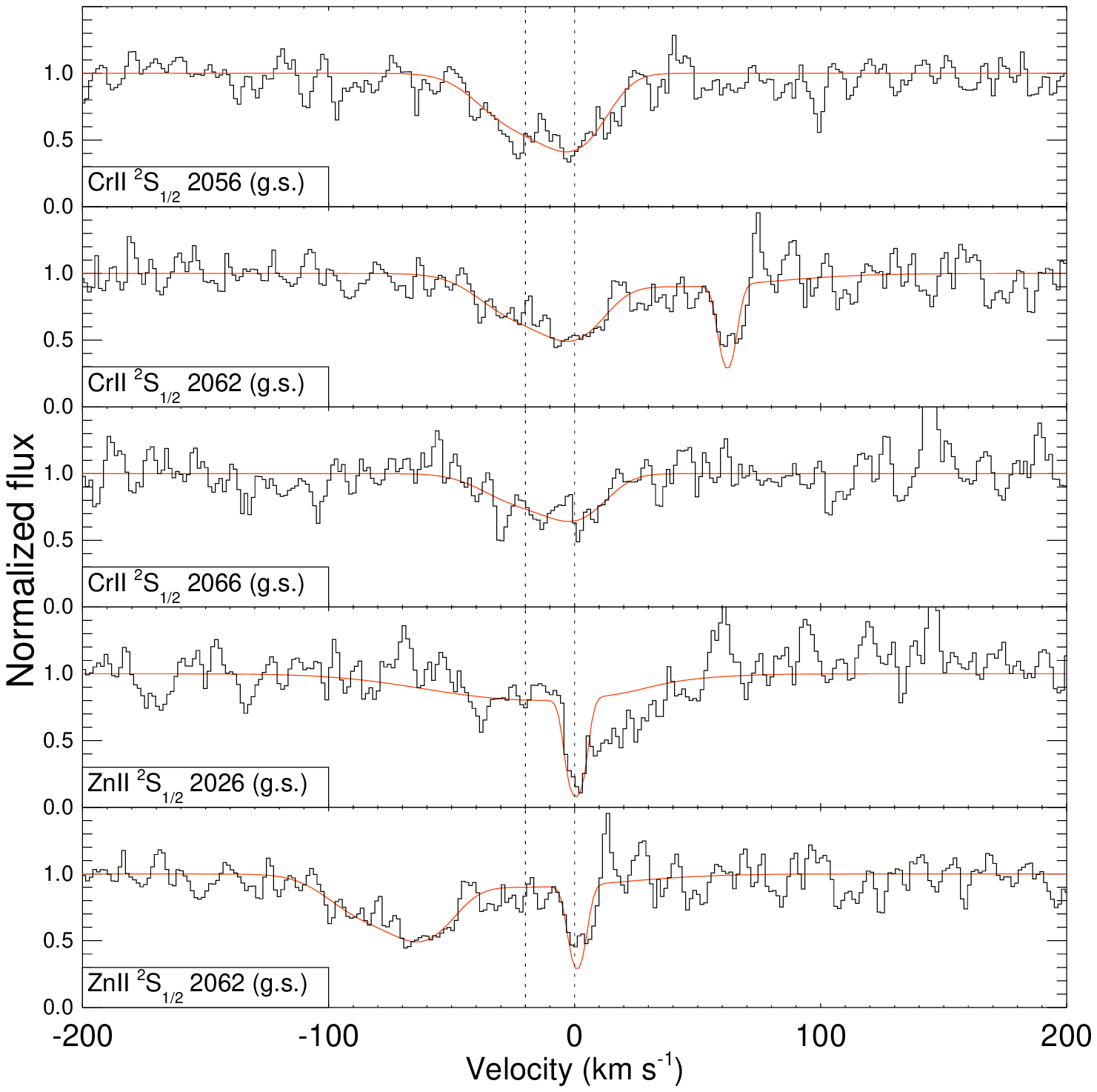}
\caption{The {Ni}{II} (top panel), {Al}{III} and {Si}{II}, (middle
  panel), {Cr}{II} and {Zn}{II} (bottom panel) absorption
  features. Solid lines represent the two Voigt components, best-fit
  model. Vertical lines identify the component velocities. The zero
  point has been arbitrarily placed at the redshift of the redmost
  component ($z=1.9683$). g.s. and n* indicate ground state and n-th
  excited transitions, respectively.}
\label{spe1}
\end{figure}

\begin{table*}
\begin{center}
\caption{Absorption line logarithmic column densities for the three components 
of the main system, derived from the UVES spectrum.}
{\footnotesize
\smallskip
\begin{tabular}{|lc|cc|cc|cc|}
\hline
\hline
Species & Observed transitions & \multicolumn{6}{c}{N (cm$^{-2}$)}      \\
\hline
{H}{I}  $^2S_{1/2}$            & Ly$\alpha$ (UVES)                               &\multicolumn{6}{c}{$21.33 \pm 0.12$}           \\
\hline
{H}{I}  $^2S_{1/2}$            & Ly$\alpha$ (FORS2)                              &\multicolumn{6}{c}{$21.11 \pm 0.10$}\\
                              & Components                                      &$20.82 \pm 0.14$ &($z_1=1.944$) &&&$20.79 \pm 0.12$ &($z_2=1.975$)        \\
\hline
\hline
Metals  &        Components              &\multicolumn{2}{c}{I ($0$ km s$^{-1}$)}&\multicolumn{2}{c}{II ($-20$ km s$^{-1}$)}&\multicolumn {2}{c}{III ($-78$ km s$^{-1}$)}\\
\hline
Species        & Observed transitions                      & N (cm$^{-2}$)& b (km s$^{-1}$) &  N (cm$^{-2}$)& b (km s$^{-1}$) &  N (cm$^{-2}$)& b (km s$^{-1}$) \\
\hline
{O}{I} $^3P_{2}$               &  $\lambda$1302                                  & $>14.34$          &SAT  & $> 14.54$       &SAT& $> 14.88$ &SAT \\
\hline
{Al}{II} $^1S_0$               &  $\lambda$1670                                  & $>12.82$          &SAT  & $> 13.26$       &SAT& $> 12.78$ &SAT \\
\hline
{Al}{III} $^2S_{1/2}$           &  $\lambda$1854, $\lambda$1862                   & $12.96 \pm 0.04$  &$12$&$  13.04\pm0.04$ &$17$& $< 12.3$ &N/A\\
\hline
{Si}{II} $^2P^{0}_{1/2}$        &  $\lambda$1260, $\lambda$1808                   & $15.44 \pm 0.03$  &$9$  &$15.08 \pm 0.18$ &$4$ & $>13.44$ &SAT \\
\hline
{Si}{II} $^2P^{0}_{3/2}$        &  $\lambda$1264, $\lambda$1816                   & $15.21 \pm 0.05$  &$9$  &$<15.0  $        &N/A & $< 15.0$ &N/A \\
\hline
{Cr}{II} $^2S_{1/2}$           &  $\lambda$2056, $\lambda$2062,  $\lambda$2066    & $13.51 \pm 0.05$  &$16$ &$13.55 \pm 0.04$ &$21$& $< 13.3$ &N/A\\
\hline
{Fe}{II} $a^6D_{9/2}$          &  $\lambda$2249,  $\lambda$2260                   & $15.09 \pm 0.03$  &$12$ &$14.95 \pm 0.04$ &$20$& $< 14.7$ &N/A\\
\hline
{Fe}{II} $a^6D_{7/2}$          &  $\lambda$1618,  $\lambda$1621                   & $13.95 \pm 0.05$  &$12$ & $ < 13.7$       &N/A & $< 13.7$ &N/A \\
\hline
{Fe}{II} $a^6D_{5/2}$          &  $\lambda$1629                                   & $13.71 \pm 0.09$  &$12$ & $ < 13.7$       &N/A & $< 13.7$ &N/A\\
\hline
{Fe}{II} $a^6D_{3/2}$          &  $\lambda$1634,  $\lambda$1636                   & $13.59 \pm 0.09$  &$12$ & $ < 13.5$       &N/A & $< 13.5$ &N/A\\
\hline
{Fe}{II} $a^4F_{9/2}$          &  $\lambda$1612,  $\lambda$1637, $\lambda$1702    & $14.19 \pm 0.05$  &$12$ &$13.76 \pm 0.07$ &$20$& $< 13.2$ &N/A\\
\hline
{Fe}{II} $a^4D_{7/2}$          &  $\lambda$1635                                   & $13.35 \pm 0.12$  &$12$ &$< 13.15$        &$20$& $< 13.15$&N/A\\
\hline
{Ni}{II} $^2D_{5/2}$           &  $\lambda$1741                                   & $13.52 \pm 0.08$  &$12$ &$13.33 \pm 0.13$ &$11$& $< 13.3$ &N/A\\
\hline
{Ni}{II} $^4F_{9/2}$           &  $\lambda$2166,  $\lambda$2217,  $\lambda$2223   & $13.57 \pm 0.02$  &$12$ &$13.27 \pm 0.04$ &$11$& $< 12.9$ &N/A\\
\hline
{Zn}{II} $^2S_{1/2}$           &  $\lambda$2026,  $\lambda$2062                   & $12.82 \pm 0.06$  &$4$  &$12.87 \pm 0.04$ &$54$& $< 12.4$ &N/A\\
\hline
\hline
\end{tabular}

}
\end{center}
\end{table*}

\begin{figure}
\centering
\includegraphics[angle=-0,width=9cm]{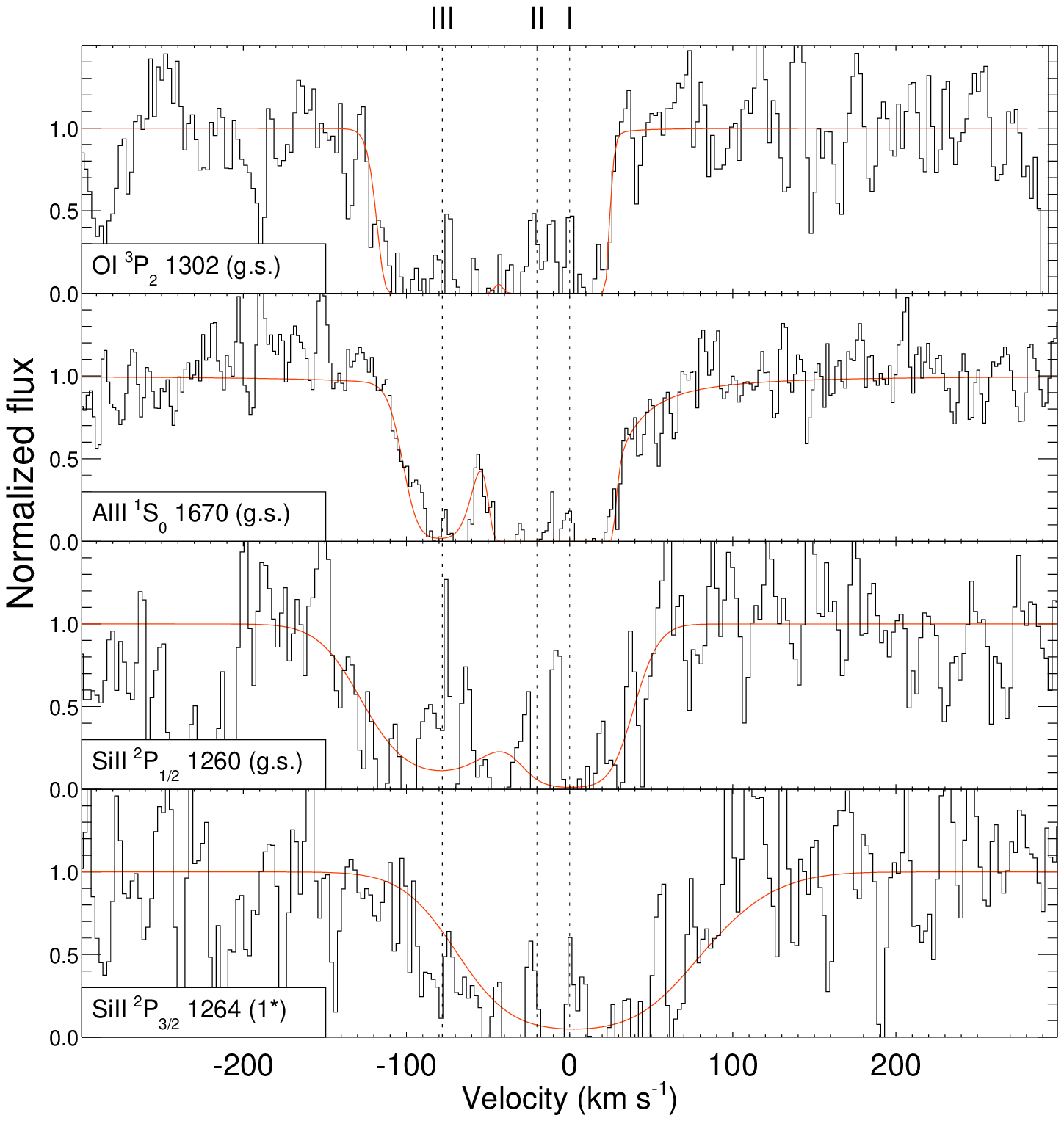}
\caption{The {O}{I}$\lambda$1302, {Al}{II}$\lambda$1670,
  {Si}{II}$\lambda$1260 ground state and {Si}{II}$\lambda$1264 fine
  structure transitions. These lines need a three Voigt component
  model to be fitted and are heavily saturated. The zero point has
  been arbitrarily placed at the redshift of the redmost component
  ($z=1.9683$). g.s. and n* indicate ground state and n-th excited
  transitions, respectively. }
\label{spe1}
\end{figure}

\subsection{Abundances}

The GRB\,081008 redshift was high enough to allow the hydrogen
Ly$\alpha$ line to enter the UVES spectral window. Unfortunately, the
UVES spectrum is extremely noisy in this region, and the derived
hydrogen column density, $\log (N_{\rm HI}/{\rm cm}^{-2}) = 21.33 \pm
0.12$ is quite uncertain. The fit is plotted in Fig. 5 (top panel)
superimposed to the smoothed UVES spectrum. The fit is particularly
poor on the wings, possibly because more than one component is needed
to model the absorption. To obtain a better estimate of the column
density, we used the FORS2 spectrum, that has a better S/N (see next
section for details). The two component fit shown in Fig. 5 (bottom
panel) gives a better representation of the data. {\bf The two
  components are centered at $z_1=1.944$ and $z_2=1.975$,
  respectively, and have column densities of $\log (N_{\rm HI}/{\rm
    cm}^{-2}) = 20.82 \pm 0.14$ and $\log (N_{\rm HI}/{\rm cm}^{-2}) =
  20.79 \pm 0.12$, respectively. The FORS2 total} column
density is $\log (N_{\rm HI}/{\rm cm}^{-2}) = 21.11 \pm 0.10$. {\bf
  This is our best fit result for $N_H$ and will be used in the
  following}. The metallicity has been derived summing all non
saturated component, excited level and ionic contributions belonging
to the same atom, dividing these values by $N_{\rm H}$ and comparing
them to the corresponding solar values given in Asplund et al. (2009).
The upper limits of component III result in an increment of the total
column densities of $< 20$\% in the worst cases, so they were not
included in the computation.
The results are listed in Table 4. Column 2 reports the total
abundance of each atom, while columns 3 and 4 report the absolute and
solar-scaled $N_X$/$N_{\rm H}$ ratios, respectively, with $X$ the
corresponding element in column 1. Lower limits are reported whenever
saturation does not allow us to securely fit the metallic column
densities (see e.g., Fig. 4, where the line profiles reach the zero
value of the normalized flux). In particular, for {O}{I} and {Al}{II}
we considered also the values of the third, saturated component, while
for {Si}{II} this has not been considered, since the fit to the
{Si}{II}$\lambda$1260 resulted in a $N$ value of component III which
is considerably lower than that of components I and II. We derived
metallicity values between $0.3$ and $0.05$ with respect to the solar
ones. We caution, however, that many transitions belonging to other
ionization states, which are commonly observed in GRB afterglow
spectra could not be taken into account, because they are outside the
UVES-dichroic-1 spectral range. In addition, the dust depletion can
prevent the observation of part of the metallic content of the
GRB\,081008 host galaxy. The reported relative abundances should then
be considered as lower limits to the true GRB\,081008
metallicity. However, some considerations on higher ionization states
are possible analyzing the FORS2 spectra, and dust content can be
investigated through the study of the depletion pattern (see sects. 5
and 6).

\begin{figure}
\centering
\includegraphics[angle=-0,width=8.5cm]{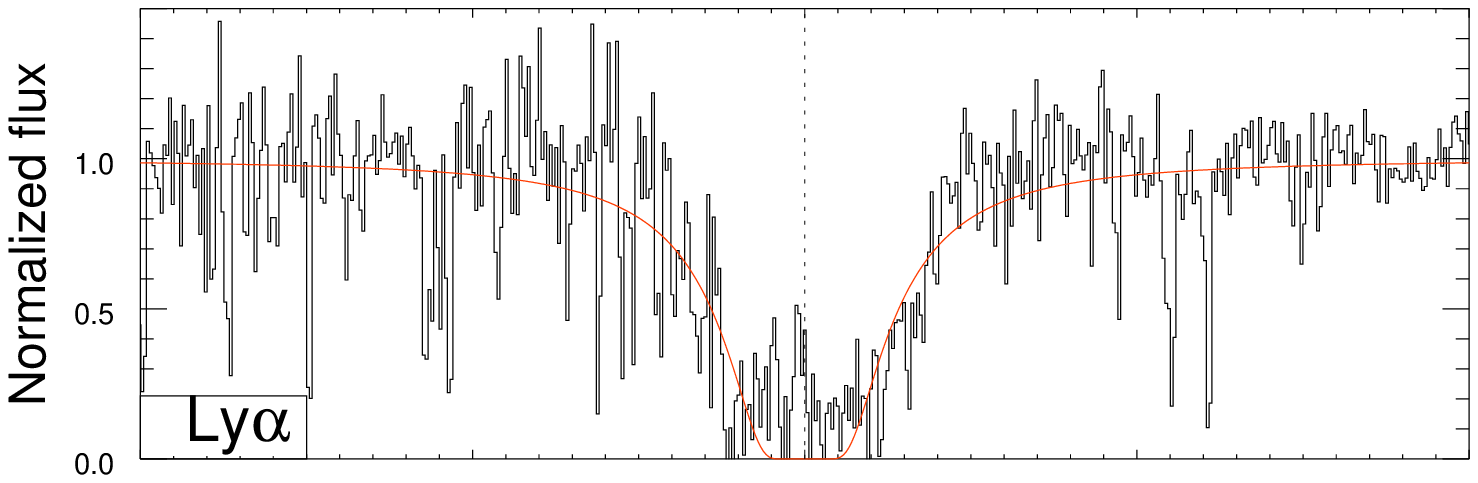}
\includegraphics[angle=-0,width=8.5cm]{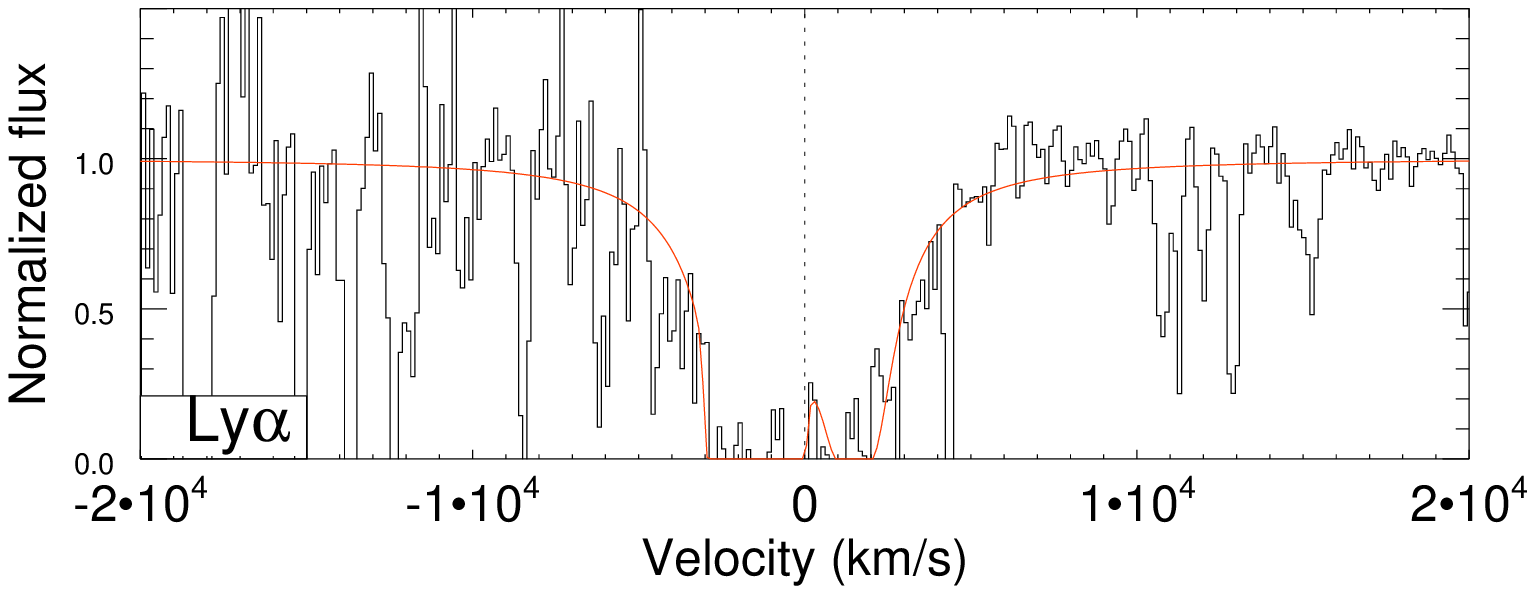}
\smallskip
\caption{The Ly$\alpha$ absorption feature at the GRB\,081008
  redshift. Top panel shows the single Voigt component, best-fit model
  for the UVES spectrum. Bottom panel shows the double component,
  best-fit model for the FORS2 spectrum. {\bf The UVES fit is poor,
    while the FORS2 one gives a more reliable description of $N_H$.} }
\label{spe1}
\end{figure}

\subsection{Excited levels}

The level structure of an atom or ion is characterized by a principal
quantum number $n$, which defines the atomic level, and by the
spin-orbit coupling (described by the quantum number $j$), which
splits these levels into fine structure sub-levels. Excited features
are routinely detected in GRB absorption spectroscopy, at the host
redshift, due to the population of both $n>1$ and/or $n=1$ fine
structure levels. GRB\, 081008 behaves the same way. In fact,
component I features the first and second fine structure levels of the
{Fe}{II} ground state ($a^6D$), the first fine structure level of
the {Si}{II} $^2P^{0}$, and the {Fe}{II} $a^4F_{9/2}$,
{Fe}{II} $a^4D_{7/2}$, {{Ni}{II}} $^4F_{9/2}$ metastable
levels (the subscript represents the spin-orbit quantum number
$j$). Moreover, the {Fe}{II} $a^4F_{7/2}$ and {{Ni}{II}}
$^4F_{9/2}$ excited states are also detected in component II (see
Table 3 for details).
 
There is conspicuous literature on the population of excited states in
GRB surrounding medium and their detection in the afterglow spectra
(see e.g. Prochaska, Chen \& Bloom 2006; Vreeswijk et al. 2007; D'Elia
et al. 2010 and references therein). There is general consensus that
these features are produced by indirect UV pumping by the afterglow,
i.e., through the population of higher levels followed by the
depopulation into the states responsible for the absorption
features. This has been demonstrated both by the detection of
variability of fine structure lines in multi-epoch spectroscopy
(Vreeswijk et al. 2007; D'Elia et al. 2009a), and through the column
density ratios of different excited levels when multiple spectra were
not available (Ledoux et al. 2009; D'Elia et al. 2009b).

Concerning GRB\,081008, the high column density of the first
metastable level of {Fe}{II} ($a^4F_{9/2}$) with respect to the fine
structure levels of the ground state can hardly be explained with a
level population distribution given by a Boltzmann function (Vreeswijk
et al. 2007), meaning that collisional excitations can be safely
rejected. The lack of multi-epoch spectroscopy does not allow us
  to completely rule out the possibility that the exciting UV flux
  come from regions of high star-formation rates and not from the
  GRB. In fact, fine structure emission lines are present in
  Lyman-break, high redshift galaxies (see Shapley et al. 2003). If
we assume that this flux comes from the GRB, we can estimate the
GRB/absorber distance, comparing observed column densities to those
predicted by a time-dependent, photo-excitation code for the time when
the spectroscopic observations were acquired.  The photo-excitation
code is that used by Vreeswijk et al. (2007) and D'Elia et
al. (2009a), to which we refer the reader for more details. Our
  equations take into account the $(4\pi)^{-1}$ correction factor to
  the flux experienced by the absorbing gas described by Vreeswijk
  (2011). We assume that the species for which we are running the
code are at the ground state before the GRB blast wave reaches the
gas. The GRB flux behavior before the UVES observation was estimated
using the data in Y10 (lightcurve and spectral index), with no
spectral variation assumed during the time interval between the burst
and our observation. We concentrate on {Fe}{II} and {Si}{II} levels
because the {Ni}{II} ground state has column densities not far from
the $90\%$ confidence level of $\log (N_{Ni}{II}/{\rm cm}^{-2}) =
13.3$, and thus the uncertainties on such values are high (Table
3). In addition, {Ni}{II} ground state is detected only through the
$\lambda 1741$ transition, because the lower oscillator strengths of
the $\lambda 1709$ and $1751$ lines prevents a detection of these
features above the $90\%$ level. The initial column densities of the
ground states were computed from the observed column densities of all
the levels of each ion, i.e., we are assuming that the species are not
excited at $t=0$. The initial values for {Fe}{II} and {Si}{II} are
$\log (N_{SiII}/{\rm cm}^{-2}) = 15.63 \pm 0.03$ and $\log
(N_{FeII}/{\rm cm}^{-2}) = 15.21 \pm 0.02$ for component I, and
$\log (N_{FeII}/{\rm cm}^{-2}) = 14.98 \pm 0.04$ for component
II. Finally, the Doppler parameter used as input of this model has
been left free to vary between $10$ and $20$ km s$^{-1}$, i.e. the
range of values that best fit the absorption features of components I
and II.

\begin{figure}
\centering
\includegraphics[angle=-0,width=9cm]{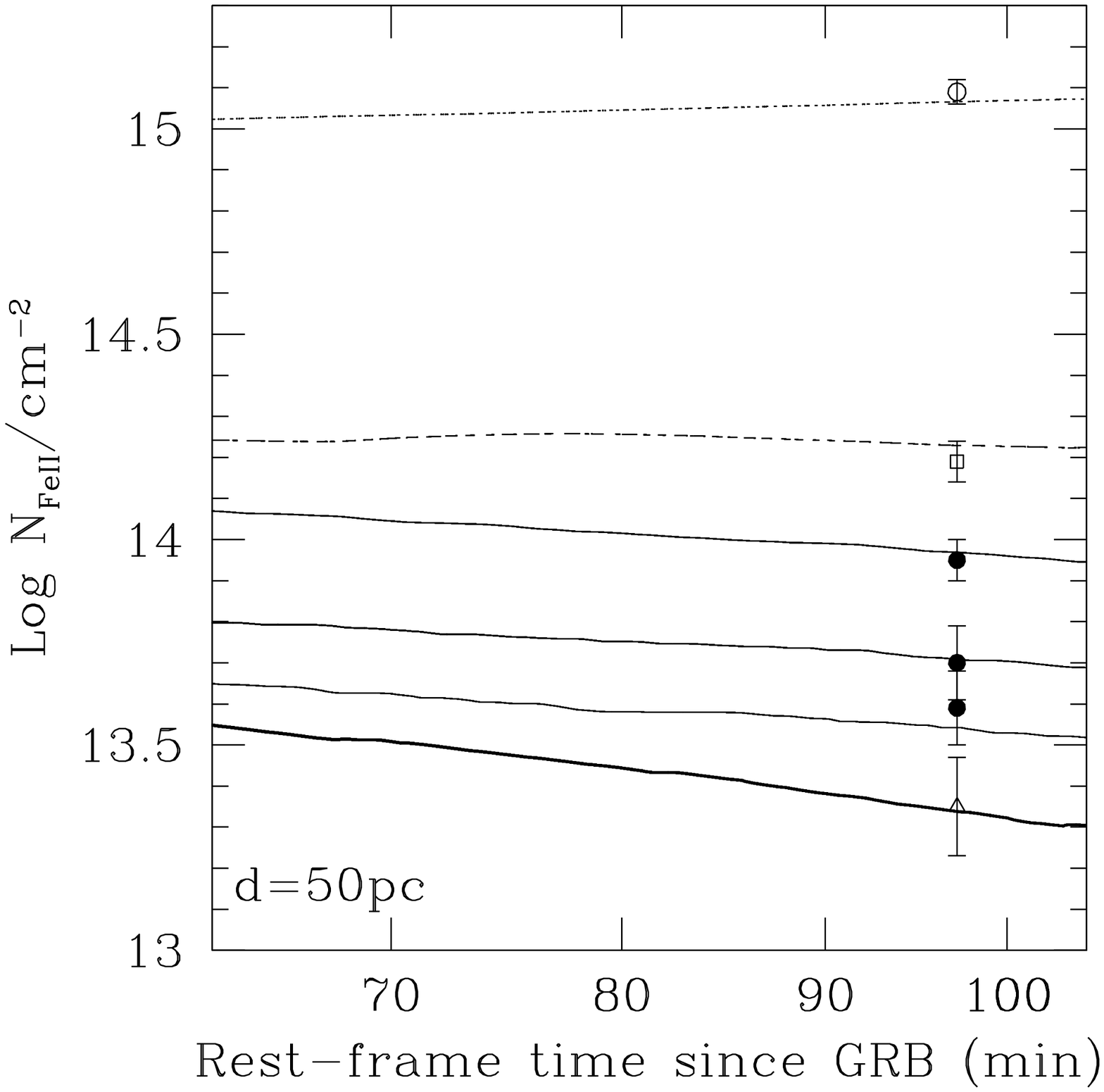}
\includegraphics[angle=-0,width=9cm]{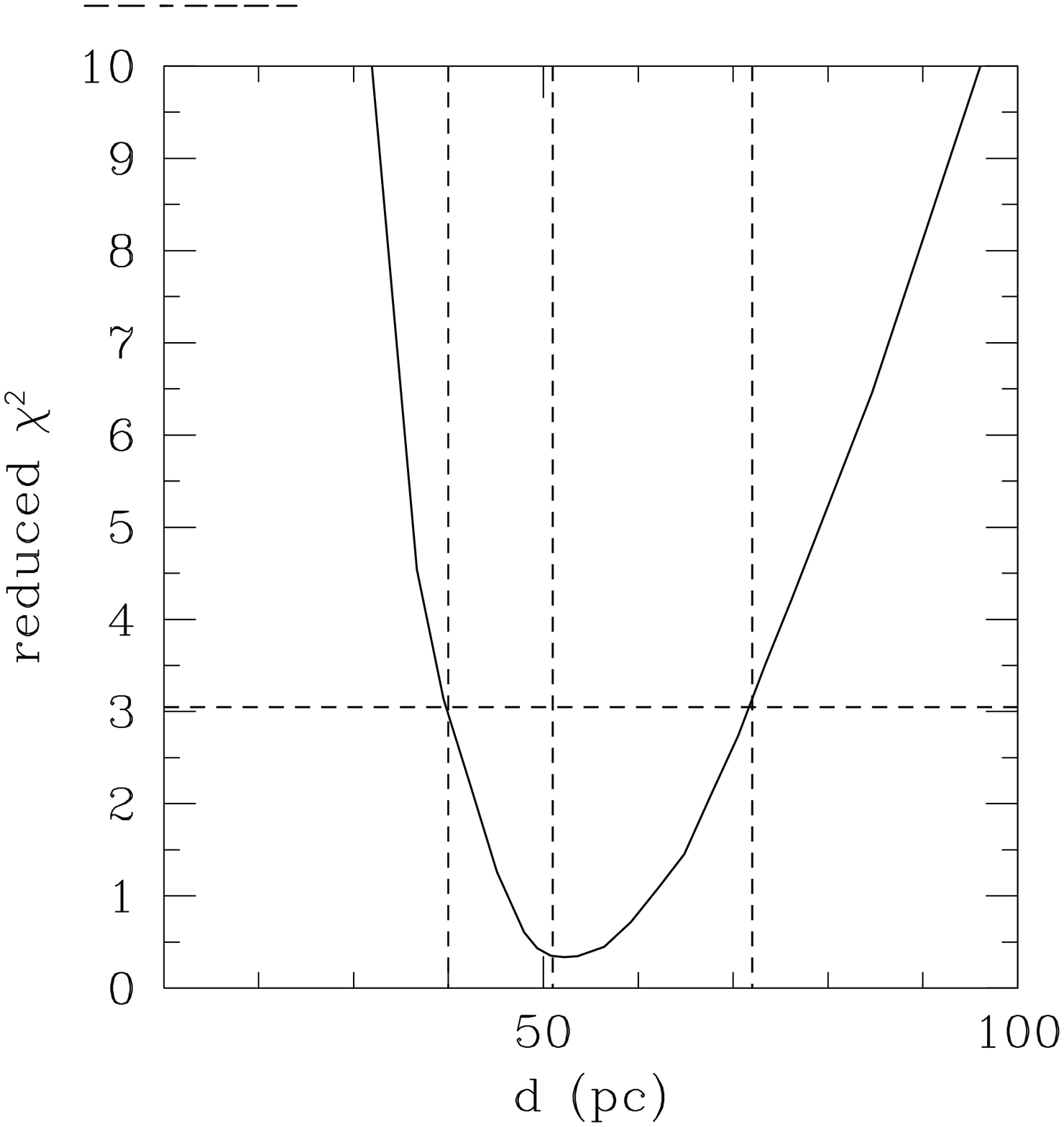}
\caption{Top panel: {Fe}{II} column densities for the ground level
  (open circle), fine structure levels of the ground state (filled
  circles), first metastable (open square) and second metastable (open
  triangle) transitions for component I in the spectrum of
  GRB\,081008. Column density predictions from our time-dependent
  photo-excitation code are also shown. They refer to the ground level
  (dotted line), fine structure level (solid lines), first and second
  excited level (dashed and thick solid lines, respectively)
  transitions, in the case of an absorber placed at $50$ pc from the
  GRB. Bottom panel: the reduced $\chi^2$ as a function of the
  distance for the model reproduced in the upper panel. Dashed lines
  indicate the best fit distance and enclose the 90\% confidence
  range.}
\label{spe1}
\end{figure}

\begin{figure}
\centering
\includegraphics[angle=-0,width=9cm]{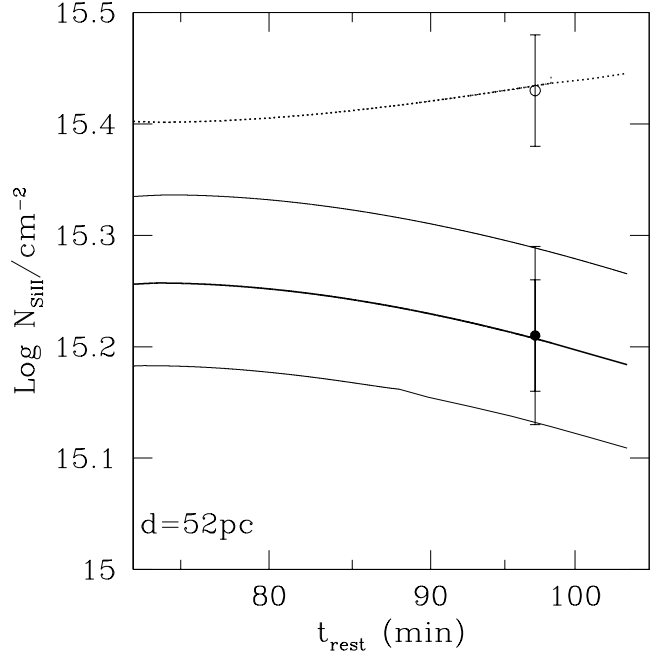}
\caption{The {Si}{II} column densities for the ground level (open
  circle) and first fine structure level (filled circle) transitions
  for component I in the spectrum of GRB\,081008.  Column density
  predictions from our time-dependent photo-excitation code are also
  shown. They refer to the ground level (dotted line) and first fine
  structure level (thick solid line) transitions, in the case of an
  absorber placed at $52$ pc from the GRB. The two thin solid lines
  display the models which enclose the fine structure level data at
  the 90\% confidence level (error bars for this transition are drawn
  both at $1\sigma$ and 90\% confidence levels).}
\label{spe1}
\end{figure}

\begin{table}
\caption{Metallicity computed from the UVES data.}
{\footnotesize
\smallskip
\begin{tabular}{|l|ccc|}
\hline
Element $X$& $\log N_X /{\rm cm}^{-2}$   &$\log N_X$/$N_{\rm H}$    & $[X/{\rm H}]$    \\
\hline
O      & $>15.12\pm0.06$  & $>-5.99\pm0.13$ & $>-2.68\pm0.11$ \\
Al     & $>13.70\pm0.04$  & $>-7.41\pm0.13$ & $>-1.86\pm0.11$ \\
Si     & $ 15.75\pm0.04$  & $ -5.32\pm0.12$ & $ -0.87\pm0.10$ \\
Cr     & $ 13.83\pm0.03$  & $ -7.28\pm0.08$ & $ -0.92\pm0.10$ \\
Fe     & $ 15.42\pm0.04$  & $ -5.69\pm0.13$ & $ -1.19\pm0.11$ \\
Ni     & $ 13.74\pm0.07$  & $ -7.37\pm0.13$ & $ -1.29\pm0.12$ \\
Zn     & $ 13.15\pm0.04$  & $ -7.96\pm0.13$ & $ -0.52\pm0.11$ \\
\hline
\end{tabular}
}
\end{table}

\begin{figure}
\centering
\includegraphics[angle=-0,width=9cm]{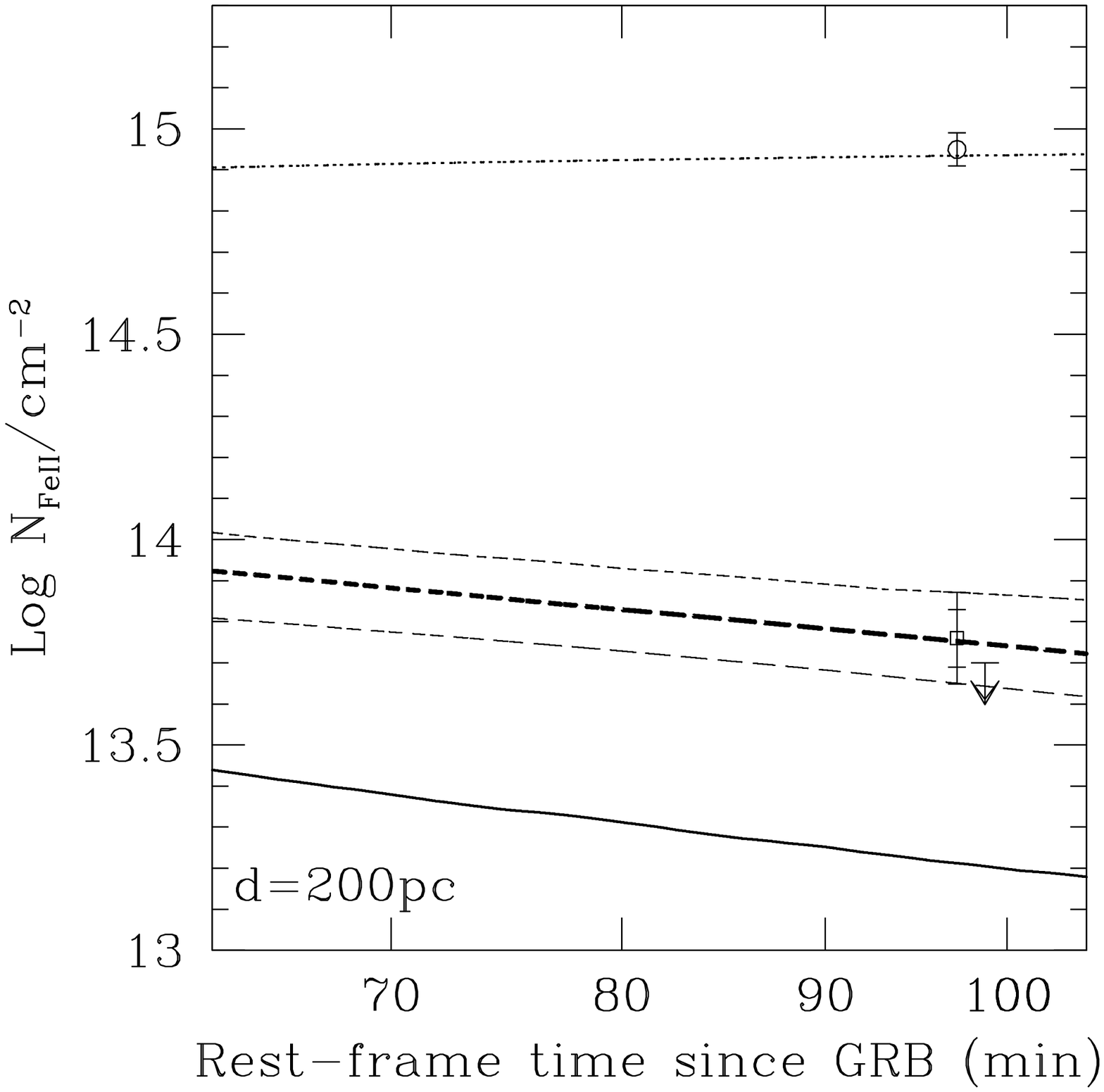}
\caption{The {Fe}{II} column densities for the ground level (open
  circle), first fine structure level (upper limit) and first excited
  level (open square) transitions for component II in the spectrum of
  GRB\,081008.  Column density predictions from our time-dependent
  photo-excitation code are also shown. They refer to the ground level
  (dotted line), first fine structure level (solid line) and
  first excited level (thick dashed line) transitions, in the case of an
  absorber placed at $200$ pc from the GRB. The two thin dashed lines
  display the models which enclose the excited level data at the 90\%
  confidence level (error bars for this transition are drawn both at
  $1\sigma$ and 90\% confidence levels).}
\label{spe1}
\end{figure}

Fig. 6 (top) shows the model that best fits the {{Fe}{II}} data,
obtained for a distance of $50$ pc and a Doppler parameter of $20$ km
s$^{-1}$. Fig. 6 (bottom) reproduces the behaviour of the reduced
$\chi^2$ as a function of the distance GRB/absorber. The distance of
component I from the GRB explosion site results $d_{I,
  FeII}=51^{+21}_{-11}$ pc at the 90\% confidence level. The same
calculation was performed using the {{Si}{II}} atomic data.  The
results are displayed in Fig. 7, and the estimated distance is
$d_{I,SiII}=52\pm 6$ pc, which is consistent with what was estimated
using the {{Fe}{II}} data.

For component II, we have much less excited transitions. Fig. 8 shows
the model which best fits the {{Fe}{II}} data and the two
theoretical curves compatible within the error bars for the
{Fe}{II} $a^4F_{9/2}$ excited level column density, which is
actually the only one with a positive detection in component II. The
resulting distance between the GRB and this absorbing component is
$d_{II, FeII} = 200^{+60}_{-80}$ pc (90\% confidence level), a
larger value than $d_{I}$, as expected given the lack of excited
transition in component II. 

\section{FORS2 spectroscopy}

In the framework of the ESO program 082.A-0755, we observed the
afterglow of GRB\,081008 also with the FORS2 low resolution
spectrograph ($R=780$), mounted on VLT/UT1. We took three spectra of
$900$ s each, starting around Oct 09 at 00:20 UT (about 4.4 hours
after the burst). We used the 600B grism, whose spectral coverage is
$330-630$ nm. The extraction of the spectra was performed within the
MIDAS environment. Wavelength and flux calibration of the three
spectra were obtained by using the helium-argon lamp and observing
spectrophotometric stars. Tab. 1 reports a summary of our FORS2
observations. We searched for variability in the W$_r$ of the FORS2
absorption lines, but we found none at the $2 \sigma$ level. This is
not surprising, since fine structure and excited lines are expected to
vary by less than $\sim 0.05$ decades (in column density) during the
acquisition time of the FORS2 spectra, which is $\sim 15$ min rest
frame (see Figs. 6-8). Since no variability is detected, we co-added
the three spectra, to improve the signal-to-noise ratio of our data,
to obtain a value of $\sim 60 - 80$ at $\lambda > 4000$\AA. The
resulting spectrum is presented in Fig. 9, together with the spectral
features identified at $z=1.97$, a redshift consistent with that
estimated using the UVES data. A list of the features detected in the
FORS2 spectrum is reported in the first column of Table 5.

\begin{figure*}
\centering
\includegraphics[width=10cm,height=20cm,angle=-90]{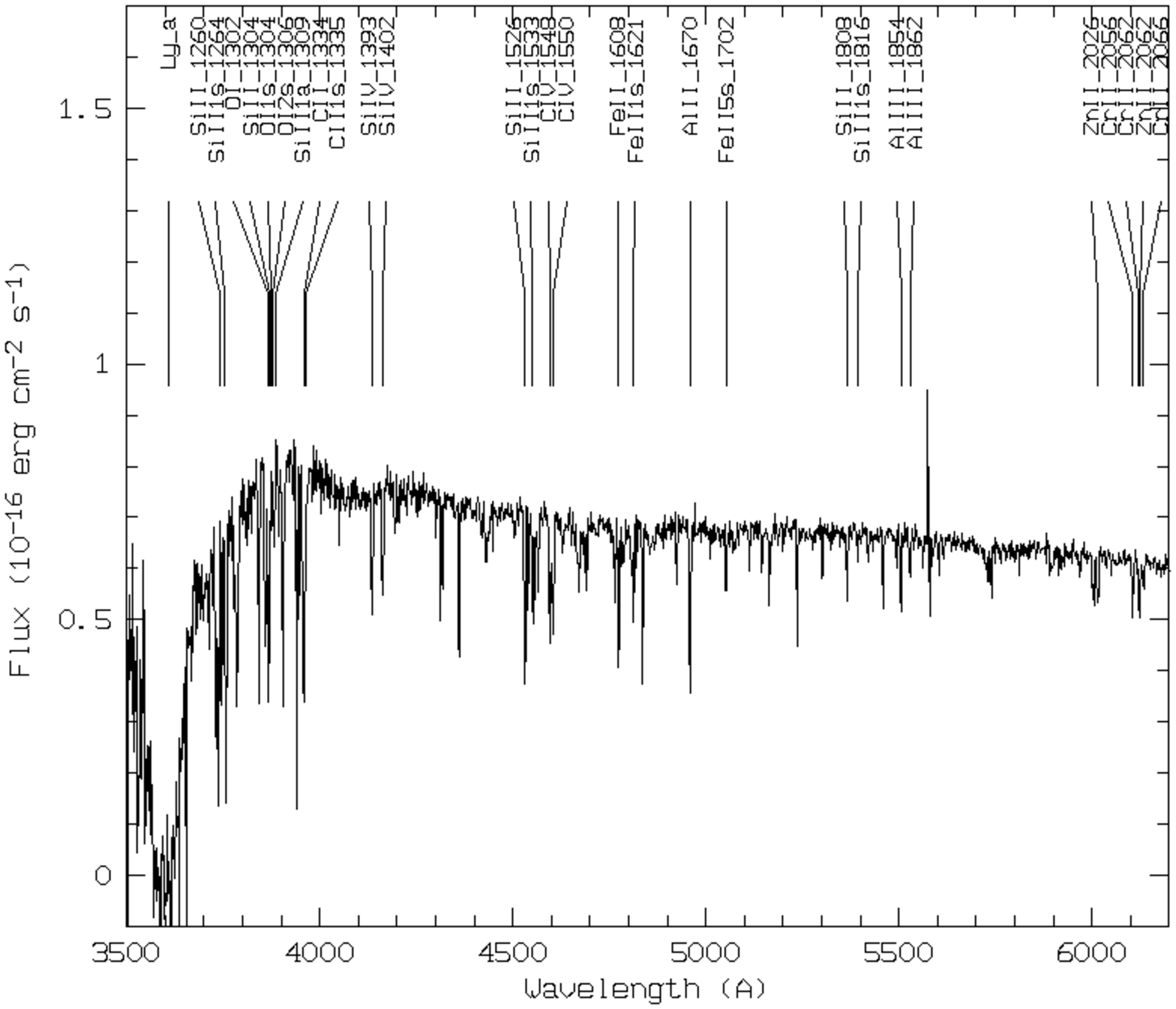}
\caption{The flux-calibrated, co-added FORS2 spectrum of the
  GRB\,081008 afterglow, together with the spectral features
  identified at $z=1.9683$.}
\label{spe1}
\end{figure*}

\begin{table}
\begin{center}
\caption{GRB081008 absorption features detected in the FORS2 spectrum, together
with their W$_r$ and column densities.
UVES data are shown for comparison.}
{\footnotesize
\smallskip
\begin{tabular}{|lcc|c|}
\hline
Transition                   & W$_r$ (\AA)$^a$     & N$^b$                     &UVES N$^b$           \\
\hline
{C}{II}$\lambda$1334     & BLEND           & -                     & -               \\
\hline
{C}{IV}$\lambda$1548     & BLEND           & -                     & -               \\
\hline
{C}{IV}$\lambda$1550     & BLEND           & -                     & -               \\
\hline
{O}{I}$\lambda$1302      & BLEND           & -                     & SAT             \\
\hline
{O}{I}$\lambda$1304    & BLEND           & -                     & -               \\
\hline
{O}{I}$\lambda$1306    & BLEND           & -                     & -               \\
\hline
{Al}{II}$\lambda$1670    & $0.74$          & $14.60^{+0.26}_{-0.20}$ & SAT             \\
\hline
{Al}{III}$\lambda$1854   & $0.30$          & $13.42 \pm 0.02$      &$13.30\pm0.03$   \\
\hline
{Al}{III}$\lambda$1862   & $0.17$          & $13.42 \pm 0.02$      &$13.30\pm0.03$   \\
\hline
{Si}{II}$\lambda$1260    & BLEND           & $15.74^{+0.15}_{-0.11}$ &$15.60 \pm 0.04$ \\
\hline
{Si}{II}$\lambda$1304    & BLEND           & $15.74^{+0.15}_{-0.11}$ &$15.60 \pm 0.04$ \\
\hline
{Si}{II}$\lambda$1526    & $0.67$          & $15.74^{+0.15}_{-0.11}$ &$15.60 \pm 0.04$ \\
\hline
{Si}{II}$\lambda$1808    & $0.25$          & $15.74^{+0.15}_{-0.11}$ &$15.60 \pm 0.04$ \\
\hline
{Si}{II} $\lambda$1264 & BLEND           & $15.17 \pm 0.01$      &$15.21 \pm 0.05$ \\
\hline
{Si}{II} $\lambda$1309 & BLEND           & $15.17 \pm 0.01$      &$15.21 \pm 0.05$ \\
\hline
{Si}{II} $\lambda$1533 & BLEND           & $15.17 \pm 0.01$      &$15.21 \pm 0.05$ \\
\hline
{Si}{II} $\lambda$1816 & $0.07$          & $15.17 \pm 0.01$      &$15.21 \pm 0.05$ \\
\hline
{Si}{IV} $\lambda$1393   & $0.47$          & -                   & -               \\
\hline
{Si}{IV} $\lambda$1402   & $0.40$          & -                   & -               \\
\hline
{Cr}{II}$\lambda$2056    & $0.27$          & $13.94 \pm 0.02$    &$13.83 \pm 0.03$ \\
\hline
{Cr}{II}$\lambda$2062    & BLEND           & $13.94 \pm 0.02$    &$13.83 \pm 0.03$ \\
\hline
{Cr}{II}$\lambda$2066    & $0.17$          & $13.94 \pm 0.02$      &$13.83 \pm 0.03$ \\
\hline
{Fe}{II}$\lambda$1608    & $0.54$          & $15.29^{+0.13}_{-0.10}$ &$15.33 \pm 0.02$ \\
\hline
{Fe}{II}$\lambda$1621  & BLEND           & -                     &$14.95 \pm 0.04$ \\
\hline
{Fe}{II}$\lambda$1629  & BLEND           & -                     &$14.95 \pm 0.04$ \\
\hline
{Fe}{II}$\lambda$1702  & $0.24$          & $14.16 \pm 0.02$      &$14.33 \pm 0.05$ \\
\hline
{Zn}{II}$\lambda$2026    & $0.24$          & $13.22 \pm 0.02$      &$13.15 \pm 0.04$ \\
\hline
{Zn}{II}$\lambda$2062    & BLEND           & $13.22 \pm 0.02$      &$13.15 \pm 0.04$ \\
\hline
\end{tabular}
}
\end{center}
$^a$ The error on W$_r$ is $0.01$\AA ($1\sigma$).

$^b$ All values of the column densities are logarithmic (in cm$^{-2}$).  
\end{table}

The Voigt fitting procedure is not adequate to compute the column
densities of metallic species in low resolution spectroscopy. In
this case, the Curve of Growth (COG) analysis (see e.g. Spitzer 1978)
must be applied. For weak absorption lines, with width $W_r < 0.1$\AA,
and for Doppler parameters $b>20$ km s$^{-1}$, $W_r$ is proportional
to the column density $N$, and virtually insensitive to the Doppler
parameter itself. For stronger lines this does not hold any more, and
the relation between $W_r$ and $N$ is described by a COG, which is a
function of $b$. In order to fit the correct COG to the data and to
estimate $b$, different transitions (with different oscillator
strengths $f$) of the same species are needed. Following Spitzer
(1978), we built up a code to perform this fit on our FORS2 data. To
test our code, we compute $W_r$ for all the UVES transitions featuring
two components, and apply our fitting program. The result of the fit
is shown in Fig. 10 (top panel), and the estimated column densities
are reported in Table 6 (errors are given at the $1\sigma$ level).

\begin{figure}
\centering
\includegraphics[angle=-0,width=9cm]{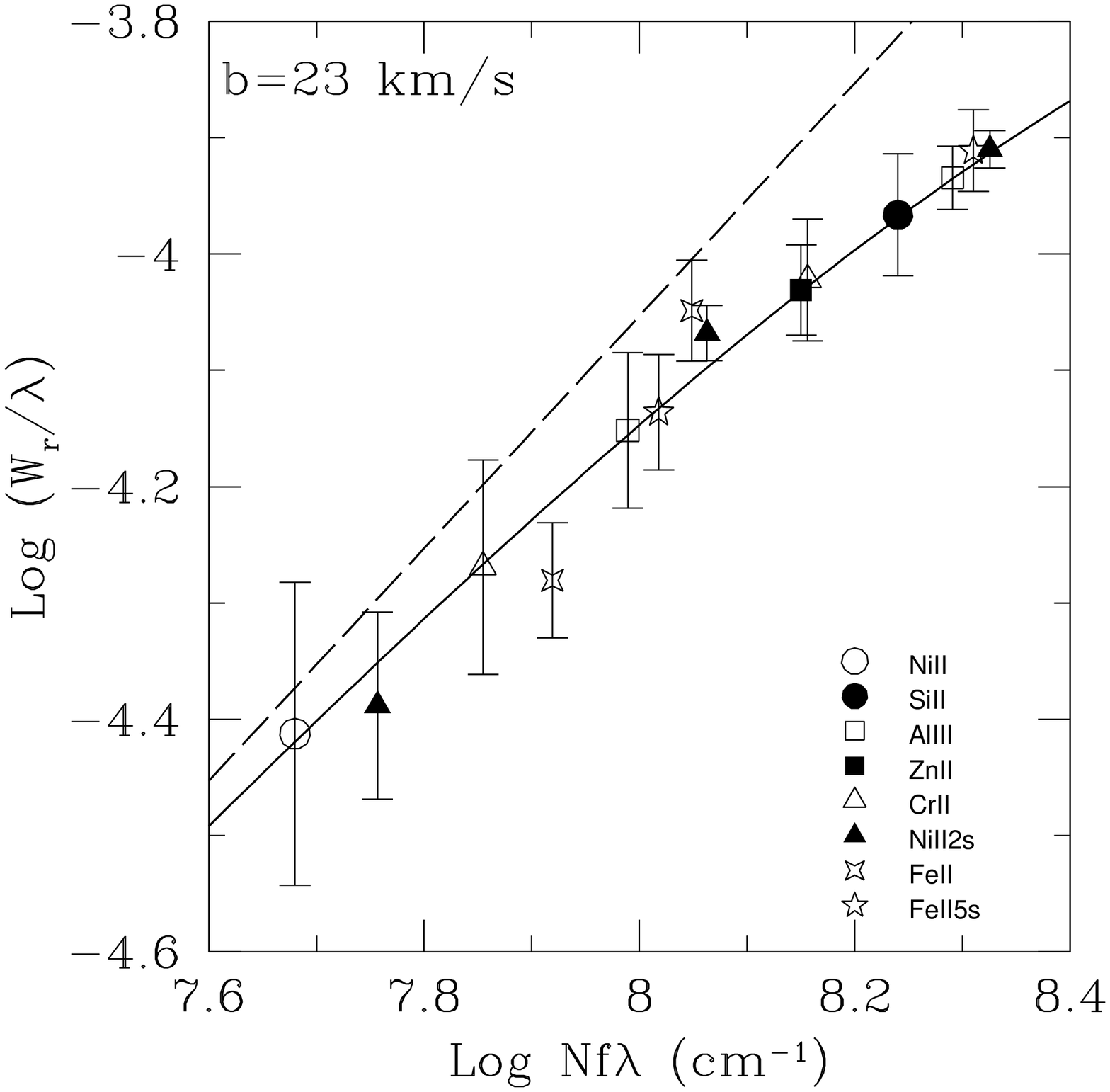}
\includegraphics[angle=-0,width=9cm]{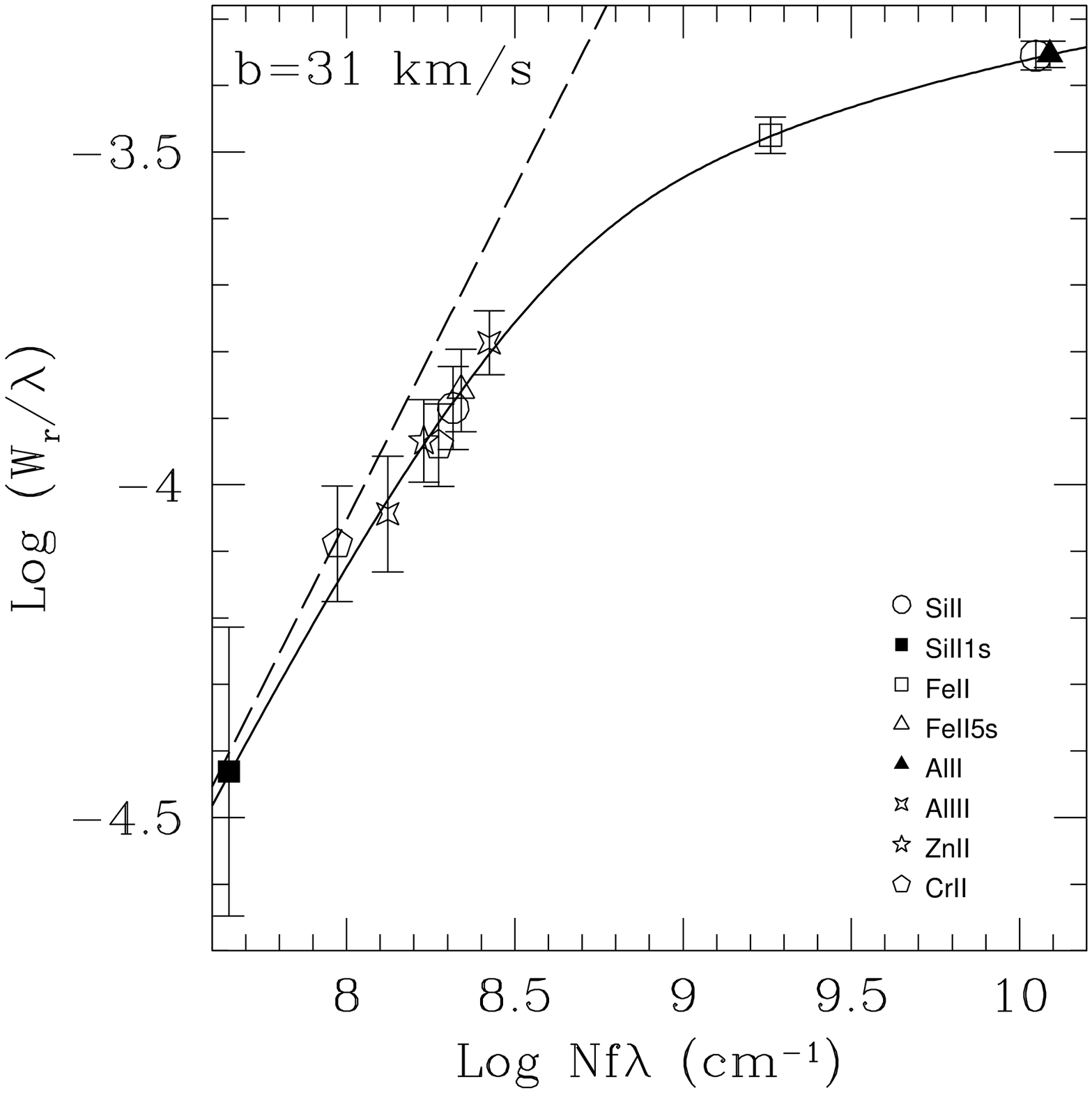}
\caption{Top panel: the COG analysis tested using the UVES species
  featuring two components. Bottom panel: COG analysis applied to the
  FORS2 lines with a measured $W_r$. Solid lines represent the best
  fit obtained using the reported $b$ values. Dashed lines show
    the $b=\infty$ curve for comparison. The COG fits component I and
    II together in both plots.}
\label{spe1}
\end{figure}

The effective Doppler parameter evaluated from the fit, $b=23$ km
s$^{-1}$, is compatible with that estimated using the line fitting
profile. To compare the column densities estimated with the two
methods, we sum for each species the contribution coming from the two
components using the line fitting method (see values in Table 3), and
report the results in Table 6. The agreement between line fitting and
COG analyses is very good: each column density is within $1\sigma$
from the corresponding value estimated using the other method. The
only exception is the {Fe}{II5s}, whose column density values
however overlap at the $2\sigma$ level.

\begin{table}
\begin{center}
\caption{Comparison between UVES column densities evalutated with the line fitting and COG methods.}
{\footnotesize
\smallskip
\begin{tabular}{|l|cc|}
\hline
Specie                   & N (COG analysis)                        &N (Line fitting)        \\
\hline
{Al}{III}                & $13.29^{+0.05}_{-0.10}$                   & $13.30 \pm 0.03$       \\
\hline
{Si}{II}                 & $15.66^{+0.06}_{-0.12}$                   & $15.60 \pm 0.03$       \\
\hline
{Cr}{II}                 & $13.83^{+0.03}_{-0.07}$                   & $13.83 \pm 0.03$       \\
\hline
{Fe}{II} (g.s.)          & $15.31^{+0.01}_{-0.07}$                   & $15.33 \pm 0.02$       \\
\hline
{Fe}{II} $a^4F_{9/2}$     & $14.13^{+0.07}_{-0.11}$                   & $14.33 \pm 0.05$       \\
\hline
{Ni}{II} (g.s.)          & $13.81^{+0.01}_{-0.03}$                   & $13.74 \pm 0.07$       \\
\hline
{Ni}{II} $a^4F_{9/2}$     & $13.73^{+0.05}_{-0.09}$                   & $13.75 \pm 0.02$       \\
\hline
{Zn}{II}                 & $13.14^{+0.05}_{-0.10}$                   & $13.15 \pm 0.04$       \\
\hline
\end{tabular}

All values are logarithmic (in cm$^{-2}$).  
}
\end{center}
\end{table}

We now apply the COG analysis to the FORS2 spectrum.  First of all, we
compute the $W_r$ from the data. As shown in Table 5, despite the
identification of nearly $30$ transitions, reliable $W_r$ can be
evaluated only for $13$ (second column of the table). This is because
the lower FORS2 resolution does not enable to separate many of these
transitions which are blended with each other. We then run the COG
code using the FORS2 $W_r$, and evaluate the corresponding column
densities. The results are shown in the third column of Table 5, while
the last column shows the UVES column densities for comparison. Errors
are again at the $1\sigma$ level, and Fig. 10 (bottom panel) shows the
graphical output of the fit.  The effective Doppler parameter
estimated ($b=31 \pm 2$ km s$^{-1}$) reproduces quite well the
combination of the values computed for component I and II ($\sim 10$
and $\sim 20$ km s$^{-1}$, respectively, separated by $\sim 20$ km
s$^{-1}$) using the line fitting method. The FORS2 and UVES spectra
give consistent column densities, with the $3\sigma$ confidence
regions overlapping in the worst cases.

\section{Conclusions and discussion}

In this paper we present high and low resolution spectroscopy of the
optical afterglow of GRB\,081008, observed using UVES and FORS2
spectrographs at the VLT $\sim 5$ hr after the trigger. We detect
several absorption features (both neutral and excited) at the common
redshift of $z=1.9683$. The spectra show that the gas absorbing the
GRB afterglow light can be described with three components identified
in this paper as I, II and III, according to their decreasing velocity
values. 

We estimated the distances between the GRB and the absorbers. We find
a distance for component I of $d_{FeII,I}=51^{+21}_{-11}$ pc and
$d_{SiII,I}=52 \pm 6$, using {{Fe}II} and {{Si}II}, respectively. The
{{Si}II} leads to a smaller uncertainty because its fine structure
level is more sensitive to the flux experienced by the absorber. Other
papers mainly use {{Fe}{II}} as distance estimator, so for a safer
comparison is better to consider our {{Fe}{II}} value. For component
II, this distance is greater, $d_{II}=200^{+60}_{-80}$ pc.  We stress
that these values are obtained assuming a three component
absorber. However, we can not exclude a higher number of components,
because our spectrum has a low S/N and a limited resolution.
Component II is far away from GRB than component I, as expected given
the lack of fine structure lines in this absorber. Component III does
not show excited levels at all, and only shows low ionization
states. Therefore, this is produced by an absorber located even
farther from the GRB, in a region which is not significantly
influenced by the prompt/afterglow emission.

Component I of GRB\,081008 is the closest to a GRB ever recorded. In
fact, for the 6 other GRBs for which the GRB/absorber distance have
been estimated, the closest components are at $d=80 - 700$ pc from the
GRB (Vreeswijk et al. 2007; D'Elia et al. 2009,a,b; Ledoux et al. 2009;
D'Elia et al. 2010; Th\"one et al. 2011). The values reported in
  literature have been corrected for the $4(\pi)^{-1/2}$ factor
  discussed by Vreeswijk (2011). This behaviour can be interpreted as
due to a dense environment close to the GRB explosion site. This high
density is possibly witnessed by the a non negligible dust amount (see
below) and by the metal content of the GRB surrounding medium. In
fact, the GRB\,081008 surroundings have the highest metallicity and
the highest abundances of, e.g., {{Fe}{II}} and {{Ni}{II}}, among this
sub-sample of GRBs. This high density in the GRB surroundings could
constitute a barrier to the GRB prompt/afterglow emission, that is not
able to strongly excite the interstellar medium up to the distances
reached by the other GRBs.

The neutral hydrogen column density is $\log (N_{\rm H, opt}/{\rm
  cm}^{-2}) = 21.11 \pm 0.10$, while that estimated from {\it Swift}
XRT data is $\log{N_{\rm H,X}}/{\rm cm}^{-2}=21.66^{+0.14}_{-0.26}$
(Campana et al. 2010). The latter value is for a solar abundance
medium. Using $N_{\rm H, opt}$ we evaluate the GRB\,081008 host
galaxy's metallicity. The values we find are in the range [X/H] $=
-1.29$ to $-0.52$ with respect to the solar abundances. This value
lies in the middle of the GRB distribution, (Savaglio 2006; Prochaska
et al. 2007; Savaglio, Glazebrook \& Le Borgne, 2009).  From X--ray
data a limit of [X/H] $> -1.83$ ($90\%$ confidence limit) can be set
assuming a solar abundance pattern and requiring that the absorbing
medium is not Thomson thick. If we set the metallicity to [X/H] =
$-0.5$, the absorbing column density in the X-rays is higher, namely,
$\log{N_{\rm H,X}}/{\rm cm}^{-2}=22.24^{+0.19}_{-0.30}$ and higher for
lower metallicities. Fynbo et al. (2009) and Campana et al. (2010)
show that in GRBs with a detectable Ly$\alpha$ feature (i.e., those at
$z>2$) $N_{\rm H,X}$ is on average a factor of $10$ higher than
$N_{\rm H,opt}$, and GRB\,081008 follows this trend. The intense GRB
flux, which ionizes the hydrogen and prevents part of it to be
optically detected, is the common explanation for this discrepancy
(Fynbo et al. 2009; Campana et al. 2010; Schady et al. 2011).

It is worth noting that observed abundances of {{Fe}{II}} and
{{Zn}{II}} are significantly different ([Fe/H]$= -1.29\pm 0.11$ and
[Zn/H]$= -0.52 \pm 0.11$). This can be ascribed to the different
refractory properties of the two elements, with the former that
preferentially tends to produce dust grains while the latter prefers
the gas phase. The comparison between these `opposite' elements can
thus provide information on the dust content in the GRB
environments. In order to be more quantitative, we derive the dust
depletion pattern for the GRB\,081008 environment, following the
method described in Savaglio (2000). We consider the four depletion
patterns observed in the Milky Way, namely, those in the warm halo
(WH), warm disk + halo (WHD), warm disk (WD) and cool disk (CD) clouds
(Savage \& Sembach 1996). We find that the best fit to our data is
given by the WH cloud pattern, with a metallicity of
$logZ_{GRB}/Z_{\odot} \sim -0.5$ and a GRB dust-to-metal ratio
comparable to that of the WH environment, e.g., $d/d_{WH}=1$
(Fig. 11). This metallicity value is consistent with our [Zn/H]
measurement. This agreement is self-consistent with the use of zinc as
a good indicator of metallicity. Since the latter quantity is linked
to the extinction (see e.g., Savaglio, Fall \& Fiore 2003) we derive
$A_V \sim 0.19$ mag along the GRB\,081008 line of sight. We check this
value by modeling the flux-calibrated FORS2 spectrum. The SED is
dominated by the Ly$\alpha$ which is difficult to model given the
high fluctuations and other absorption lines. Anyway, the inferred $A_V$
value is low and compatible with that evaluated from the dust
depletion.

\begin{figure}
\centering
\includegraphics[angle=-0,width=10cm]{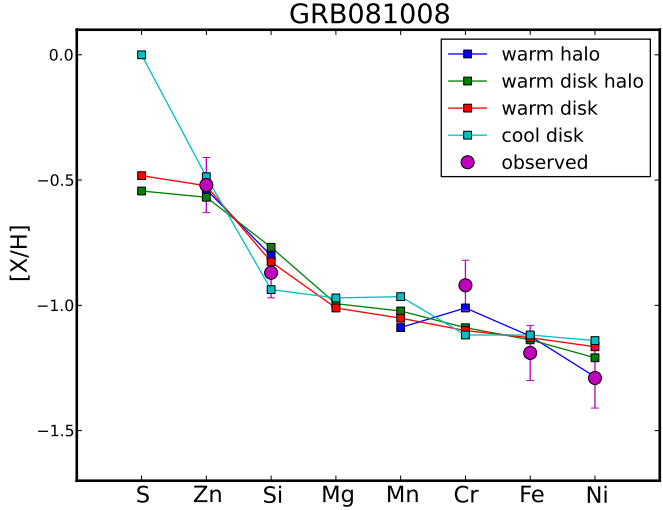}
\caption{Depletion patterns in the absorbing gas of
  GRB\,081008. Filled squares are taken from average gas-phase
  abundance measurements in warm halo (blue), warm disk + halo
  (green), warm disk (red) and cool disk (cyan) clouds of the Milky
  Way (Savage \& Sembach 1996). Filled circles represent our data
  points, which are best fitted by the warm halo cloud pattern.}
\label{spe1}
\end{figure}

Another hint of dust is the non detection of {{Fe}{II}} in the third
component. The {{Fe}{II}} column densities in components I and II are
very similar, and this lets us believe that in component III the iron
is present as well, but in the dust form. The higher presence of dust
in components far away from the GRB has already been pointed out by
D'Elia et al. (2007).  They report a possible presence of dust in
component III of GRB\,050730, while the closer component II (featuring
{{Fe}{II}} fine structure lines) shows more iron in the gas state. The
detection of more dust far away from the GRB can be explained since
dust grains containing iron tends to be efficiently destroyed during a
blast wave occurring after a GRB explosion (Perna, Lazzati \& Fiore
2003).

The analysis of the FORS2 spectra extends our surveyed wavelength
range, allowing the detection of higher ionization species, such as
{{C}{IV}} and {{Si}{IV}}. Anyway, line profiles of high and low
ionization species rarely match in redshift space and often if they
do, it is because the line blending cannot be resolved in a spectrum,
regardless of resolution and S/N.

We stress that the availability of simultaneous high and low
resolution spectra of a GRB afterglow is an extremely rare event. In
this context, the comparison of the column densities obtained fitting
the line profile of a high resolution spectrum with that estimated by
the Curve of Growth analysis applied to a low resolution one could be
extremely important. In fact, this can help to determine a range of
column densities for which it is safe to apply the Curve of Growth
analysis when high resolution data are missing. This is because high
column densities can result in the saturation effect, a problem that
is difficult to address using low resolution spectra only (see
e.g. Penprase et al. 2010).  Prochaska (2006) widely discuss the
limits and perils of the COG analysis applied to low resolution
data. They find that this kind of analysis tends to underestimate the
column densities of the absorbing species. This is because strong
transitions drive the COG fit since the relative error associated to
their W$_r$ is smaller than that for weak ones. Nevertheless, strong
transitions are more affected by saturation, and in order to match
their observed column densities, the COG fit is forced towards high
values of the effective Doppler parameter. High resolution data often
show that the main contribution to the column density of strong
transitions comes from one narrow component. On the other hand, the
main contribution to the W$_r$ comes from other components which
account for a small fraction of the column density. These inferred
high values for the effective Doppler parameter are thus mimicking a
more complex situation, with the result of underestimating the real
column densities. For what concerns GRB\,081008 the UVES observations
show no or just mild saturation even for the strongest transitions,
and the two main components give a similar contribution to the total
column densities. This is the reason why there is a good agreement
between COG analysis of low resolution data and line fitting analysis
of high resolution ones for this particular GRB (within $3\sigma$ in
the worst cases).


Finally, we detect two weak intervening systems in our spectra. The
first one is a {{C}{IV}} absorber in the FORS2 spectrum at $z =1.78$,
and the second one is a {{Mg}{II}} system in the UVES spectrum at
$z=1.286$. This last system has $W_r({{Mg}{II}\lambda 2796})=0.3$\AA,
the detection limit being $0.1$\AA\, at the $2\sigma$ confidence
level. The redshift path analyzed for {{Mg}{II}} is $z=0.18-0.38$ and
$z=0.71-1.43$ for the UVES spectrum, and $z=0.36-1.21$ for the FORS2
one.

\section*{Acknowledgments}
We thank an anonymous referee for a deep and critical reading of the
paper, which strongly increased its quality. This work was partially
supported by ASI (I/l/011/07/0).

\end{document}